\documentclass[12pt]{iopart}
\pdfoutput=1

\usepackage{siunitx}
\usepackage{graphicx}

\usepackage{gensymb}
\usepackage{amssymb}
\usepackage{revsymb}	

\newcommand{\ve}[1]{\mathbf{#1}}
\newcommand{\E}{\mathcal{E}}
\newcommand{\0}{\phantom{0}}

\newcommand{\px}{p_\textup{x}}
\newcommand{\py}{p_\textup{y}}
\newcommand{\pz}{p_\textup{z}}
\newcommand{\Ex}{E_\textup{x}}
\newcommand{\Ey}{E_\textup{y}}

\newcommand{\Bz}{B_\textup{z}}
\newcommand{\me}{m_\textup{e}}

\newcommand{\ES}{E_\textup{S}}
\newcommand{\chie}{\chi_\textup{e}}
\newcommand{\Egamma}{\E_\gamma}
\newcommand{\kB}{k_\textup{B}}
\newcommand{\Tgamma}{T_\gamma}
\newcommand{\Te}{T_\textup{e}}

\newcommand{\etas}[1]{\eta_{_\textup{#1}}}

\newcommand{\fullfigure}{0.5\columnwidth}
\newcommand{\fullfigurepng}{0.8\columnwidth}
\newcommand{\halffigure}{0.4\columnwidth}

\newcommand{\tgtfigureheight}{150pt}
\newcommand{\figurespacing}{\hspace{0.17\columnwidth}}

\begin{document}

\title{Inverse Compton scattering from solid targets irradiated by ultra-short laser 
pulses in the \SI{e22}{} -- \SI{e23}{W/cm^2} regime}

\author{Ji\v{r}\'{i} Vysko\v{c}il$^{1}$, Evgeny Gelfer$^{2}$ and Ond\v{r}ej Klimo$^{1,2}$}

\address{$^1$ Faculty of Nuclear Sciences and Physical Engineering, Czech Technical University in Prague, B\v{r}ehov\'{a} 7, 11519, Prague, Czech Republic}
\address{$^2$ Institute of Physics AS CR, ELI-Beamlines project, Za Radnic\'{i} 835, 25241 Doln\'{i} B\v{r}e\v{z}any, Czech Republic}

\date{\today}

\begin{abstract}
Emission of high energy gamma rays via the non-linear inverse Compton scattering process (ICS) in 
interactions of ultra-intense laser pulses with thin solid foils is studied using particle-in-cell 
simulations. It is shown that the angular distribution of the ICS photons has a forward-oriented 
two-directional structure centred at an angle $\vartheta = \pm\ang{30}$, a value predicted by a theoretical 
model based on a standing wave approximation to the electromagnetic field in front of the target, which only 
increases at the highest intensities due to faster hole boring, which renders the approximation invalid. The 
conversion efficiency is shown to exhibit a super-linear increase with the driving pulse intensity. In 
comparison to emission via electron-nucleus bremsstrahlung, it is shown that the higher absorption, further 
enhanced by faster hole boring, in the targets with lower atomic number strongly favours the ICS process.
\end{abstract}


\submitto{\PPCF}
\maketitle


\section{Introduction}

Next generation high-power laser systems are expected to routinely reach intensities in the $I \approx \num{e22} -
\SI{e23}{W/cm^2}$ region \cite{WeberP3installationhighenergy2017, HernandezGomezVulcan10PW2010,
    ZouDesigncurrentprogress2015, DansonPetawattclasslasers2015}. In a configuration where such an intense pulse 
    interacts
with a solid target, gamma rays will be generated mostly by the processes of electron-nucleus
bremsstrahlung~\cite{BernsteinRayProductionLaser1970}, and by radiation reaction effects including non-linear inverse
Compton scattering (ICS) \cite{RitusQuantumeffectsinteraction1985, LauNonlinearThomsonscattering2003}, where the fast
electrons scatter on the high field of the laser pulse itself \cite{NakamuraHighPowergRayFlash2012}. In this 
paper, we
present a study of the latter process, relevant especially at the higher end of the considered intensity range, where
the radiation has to be treated in the context of quantum electrodynamics (QED), with the further outlook of even higher
intensities which would exhibit additional important effects such as the creation of electron-positron pairs and QED
cascades \cite{GuGammaphotonselectronpositron2019, BellPossibilityProlificPair2008,
    BulanovElectromagneticcascadehighenergy2013, DiPiazzaExtremelyhighintensitylaser2012,
    RidgersDenseElectronPositronPlasmas2012, BradyLaserAbsorptionRelativistically2012, 
    ZhidkovRadiationDampingEffects2002,
    TaPhuocAllopticalComptongammaray2012}.

The non-linear multi-photon nature of the ICS process requires the presence of fast electrons and high fields. In the
context of laser-plasma interactions, it has been observed in various configurations where the laser pulse interacts
with an accelerated electron beam. Early observations \cite{EnglertSecondharmonicphotonsinteraction1983,
    BulaObservationNonlinearEffects1996, ChenExperimentalobservationrelativistic1998,
    SchwoererThomsonBackscatteredRaysLaserAccelerated2006, MalkaRelativisticelectrongeneration2002,
    ChenMeVEnergyRaysInverse2013} 
of multi-photon scattering on fast electrons were limited to the regime of low energy of the emitted photons,
$\hbar\omega_\gamma \ll \me c^2$, where $\hbar$ is the reduced Planck constant, $\omega_\gamma$ the photon's angular 
frequency, $\me$ the electron mass, and $c$ the speed of light, which is commonly called non-linear Thomson scattering 
as opposed to (non-linear) inverse Compton scattering where $\hbar\omega_\gamma \gg \me c^2$ 
\cite{LauNonlinearThomsonscattering2003}. These were followed by observations of the ICS interaction in the non-quantum 
regime in experiments with laser wakefield accelerated electrons and a counter-propagating laser pulse with the gamma 
ray energies of $\Egamma = 6 - \SI{18}{MeV}$ \cite{SarriUltrahighBrillianceMultiMeV2014}, and $\Egamma > \SI{20}{MeV}$ 
\cite{YanHighordermultiphotonThomson2017}, though the authors stick to calling the interaction the non-linear Thomson 
process in order to highlight that the quantum effects are still negligible in this regime. The energies high enough to 
probe the quantum nature of the interaction, as opposed to the classical radiation reaction approximation, were not 
reached until 2018 when a landmark experiment by Cole et al.\ \cite{ColeExperimentalEvidenceRadiation2018}, performed 
at the Astra Gemini laser, presented evidence of radiation reaction in the collision of an ultra-relativistic 
$\E_\textup{e} > \SI{500}{MeV}$ electron beam generated by laser-wakefield acceleration with an intense $I = 
\SI{1.3e21}{W/cm^2}, a_0 = 25$ laser pulse. The energy loss in the post-collision electron spectrum was correlated with 
the detected $\Egamma > \SI{30}{MeV}$ gamma ray signal, and was found to be consistent with a quantum  description of 
radiation reaction. A further experiment \cite{PoderExperimentalSignaturesQuantum2018} with a $I=\SI{4e20}{W/cm^2}$ 
pulse provided additional signatures of quantum effects in the electron dynamics in the external laser field, 
potentially showing departures from the constant cross field approximation.

Unlike the experiments where an intense laser pulse interacts with a solitary electron beam, the hot electrons 
participating in ICS in the laser-solid interactions studied in this paper are self-generated at the front side of the 
target due to the absorption \cite{MalkaExperimentalConfirmationPonderomotiveForce1996, 
PukhovThreeDimensionalSimulationsIon2001} of a portion of the energy of the same pulse with which they 
immediately 
interact giving out high-energy gamma rays. By means of Particle-in-Cell simulations using the code 
EPOCH \cite{ArberContemporaryparticleincellapproach2015}, we study the ICS emission from thin foils as a
function of the laser pulse intensity, describe its energy spectrum and angular distribution, and present a simplified 
standing-wave model that explains some of the emission's prominent features. Additionally, we examine the effect of 
target material, and compare the ICS emission to bremsstrahlung, which we studied in our previous paper 
\cite{VyskocilSimulationsbremsstrahlungemission2018} under the same conditions.

The paper is organized as follows. \Sref{theory} summarizes the essential theoretical background, and
\sref{simulation-setup} describes the PIC simulation setup. \Sref{results} presents the results, in 
particular the simulated ICS photon energy spectra, the simplified standing wave model and its comparison to 
the PIC simulations, the description of electron dynamics at the front side of the target, the predicted 
emission angle of the ICS photons and the angular distribution obtained from the PIC simulations, the 
efficiency of conversion of the driving laser pulse energy into that of the ICS photons, and a comparison of 
ICS to bremsstrahlung emission. \Sref{conclusions} summarizes our conclusions.

\section{Gamma ray emission by inverse Compton scattering}\label{theory}
The ICS radiation is in fact not emitted continuously. Individual photons are emitted as the electron loses 
energy due to its interaction with the strong field. To characterize this interaction, taking into account 
the discontinuous nature of the process, a parameter $\chie$ is introduced 
\cite{RitusQuantumeffectsinteraction1985, BulanovElectromagneticcascadehighenergy2013, 
KirkPairproductioncounterpropagating2009}:
\begin{equation}
\label{chi-e}
\chie = \frac{1}{\ES}\sqrt{
    \left(\gamma\ve{E} + \frac{\ve{p}\times\ve{B}}{\me}\right)^2 
    - \left(\frac{\ve{p}\cdot\ve{E}}{\me c}\right)^2} \\
\end{equation}
where $\ve{E}$ is the electric field, $\ve{B}$ is the magnetic field, $\ve{p}$ is 
the electron momentum, $\gamma = 1/\sqrt{1-v^2/c^2}$ is the relativistic Lorentz 
factor of the electron, and $\ES$ is the ``Sauter-Schwinger'' field \cite{SauterUeberVerhaltenElektrons1931, 
SchwingerGaugeInvarianceVacuum1951}, a critical field with enough strength to be able to perform $\me c^2$ 
work over the electron Compton length $\lambdabar_\textup{C} = \hbar/\me c$ 
\cite{BulanovElectromagneticcascadehighenergy2013}, $\ES = \me^2 c^3/e\hbar = \SI{1.32e16}{V/cm}$. Regarding 
the emission of gamma rays, the value of $\chie$ indicates the strength of the radiation process, roughly 
separating the classical regime $\chie \ll 1$ with continuous emission, and the quantum regime, where 
$\chie$ approaches unity and the process must be treated as a discontinuous emission of photon quanta 
\cite{ShenEnergyStragglingRadiation1972, RidgersDenseElectronPositronPlasmas2012}.

The intensity of the gamma radiation emitted by the electron can be expressed in the limits of $\chie \ll 1$ 
or $\chie \gg 1$ respectively as
\numparts
\begin{eqnarray}\label{rad-int-low}
I_\textup{rad}^{<} &= \frac{e^2\me^2}{6\pi} \chie^2 ( 1 - c_1 \chie + c_2\chie^2 - \ldots), \\
\label{rad-int-high}    I_\textup{rad}^{>} &= c_3\frac{e^2 \me^2}{6\pi} \chie^{2/3}
( 1 - 
c_4 \chie^{-2/3} + 
c_5 \chie^{-4/3}  - \ldots)
\end{eqnarray}
\endnumparts
where $e$ is the elementary charge, and $c_1, \ldots, c_5$ are constants 
\cite{NikishovQuantumprocessesfield1964}. We can then give a rough estimate of 
the extreme limits for radiation intensity. At very small $\chie$, we can only keep the unit term in the 
brackets of equation \eref{rad-int-low}, and the radiation intensity behaves as $I_\textup{rad} \sim 
\chie^2$, while at very large $\chie$, those terms in the brackets of equation \eref{rad-int-high} which are 
inversely proportional to $\chie$ raised to some positive power can be neglected, and the radiation 
intensity then behaves as $I_\textup{rad} \sim \chie^{2/3}$.

Previous equations show that in order to generate large amounts of high energy gamma rays, one needs to 
employ a high field, hot electrons, or both. The strength of the laser pulse can be expressed in terms of 
the normalized amplitude of the vector potential
\begin{equation}\label{normalized-potential}
a_0 = \frac{eE_0}{\me \omega c} \approx (7.3 \times 10^{-19} (\lambda [\si{\um}])^2 I 
[\si{W\,cm^{-2}}])^{1/2},
\end{equation}
where $E_0$ is the peak amplitude of the electric field of the laser pulse, $\omega$ its angular frequency, 
and $\lambda$ its wavelength. The temperature of the hot electrons pulled out of a solid target by a pulse 
in the non-linear relativistic regime is given by 
\begin{equation} \label{wilks-scaling}
\Te = \me c^2 (\gamma - 1),
\end{equation}
the relativistic $\gamma$ factor can be, in laser-solid interactions, approximated from the ponderomotive 
scaling \cite{WilksAbsorptionultraintenselaser1992} in the case of linear polarization as
\begin{equation}
\gamma = \sqrt{1 + \frac{a_0^2}{2}},
\end{equation}
For high values of $a_0$, this leads to a linear dependence $\Te \sim a_0$.

\section{Simulation setup}\label{simulation-setup}

Simulations were done in 2D in a $x \in (-15, 15)\;\si{\um}$, and $y \in (-20, 20)\;\si{\um}$ box with a 
cell size of $10 \times \SI{10}{nm}$. A normally incident laser pulse polarized in the simulation plane with 
a wavelength $\lambda = \SI{1}{\um}$, and a Gaussian spatial and temporal profile with a FWHM duration of 
$\tau = \SI{30}{fs}$, was propagating along the $x$ axis, and focused to a $w = \SI{3}{\um}$ spot at the 
front side of the target placed at $x = 0$. The laser pulse was emitted from the $x = \SI{-15}{\um}$ 
boundary at the start of the simulation $t = 0$, at an angle of $\vartheta_\mathrm{L} = \ang{0}$ with its 
peak intensity crossing the boundary at $t=\SI{60}{fs}$. The target was composed of a fully ionized CH 
plasma with electron density $n_\textup{e} = 289 n_\textup{c}$, where $n_\textup{c} = \epsilon_0 
m_\textup{e} \omega^2 / e^2$ is the plasma critical density which is a function of the angular frequency 
$\omega$ of the laser pulse, with $\epsilon_0$ being the permittivity of free space. For a $\lambda = 
\SI{1}{\um}$ laser pulse, the value of $n_\textup{c} = \SI{1.1e21}{cm^{-3}}$. Parameter scans were performed 
for six laser pulse intensities between $I = \SI{3e21}{W/cm^2}$, and $I = \SI{e23}{W/cm^2}$. The normalized 
potential corresponding to the intensities in the simulations ranges from $a_0 = 47$ to $a_0 = 270$. Two 
additional materials, Al, and Au, were examined in order to compare these results to our previous work 
\cite{VyskocilSimulationsbremsstrahlungemission2018}, which also describes their respective simulated parameters.

The simulations used a second order FDTD Maxwell solver \cite{YeeNumericalsolutioninitial1966a}, and a relativistic 
Boris pusher \cite{Boris1970}. To limit noise and numerical heating \cite{ArberContemporaryparticleincellapproach2015}, 
the simulations included a current smoothing algorithm and third order particle weighting. All boundary conditions were 
absorbing for radiation and thermalizing for particles. The radiation reaction effects were calculated EPOCH's Monte 
Carlo algorithm \cite{DuclousMonteCarlocalculations2010}, and bremsstrahlung 
\cite{VyskocilSimulationsbremsstrahlungemission2018} was taken into account in order to obtain a self-consistent 
results. This paper only uses the photons with $\E_\gamma > \SI{1}{MeV}$ in all subsequent analysis.

\begin{figure}
    \figurespacing\includegraphics[height=\tgtfigureheight]{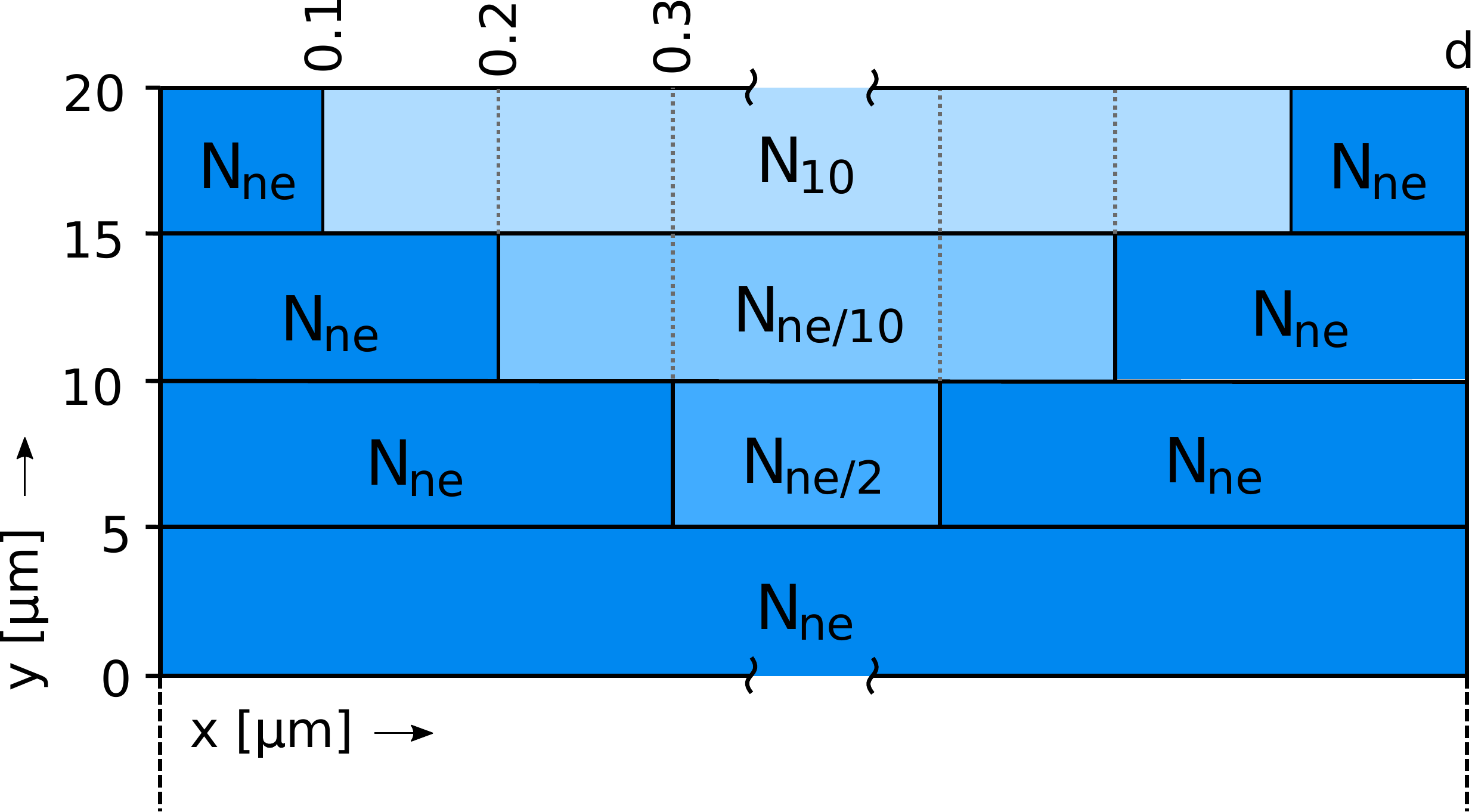}     
    \caption{\label{tgt-structure}Schematic representation of the splitting of the simulated 
    target with the width $d$
        into regions containing different number of computational macro-particles per cell. The 
        figure only shows the top
        half $y>0$ of the target, with the bottom half being symmetric with respect to the $x$ 
        axis. The axes in the figure
        are not to scale.}
\end{figure}

\Table{\label{tgt-structure-tbl}Number of macro-particles in different regions of the simulated target. 
 	The first
    column lists the designation $N_\mathrm{ppc}$ of a computational region, which is equal to the 
    number of electron
    macro-particles per cell in that region. The second column lists the value of $N_\mathrm{ppc}$, 
    i.e.\ the number of
    macro-particles per cell. The third column lists the number of ion macro-particles per cell 
    either as an absolute
    value, or with respect to the number of electron macro-particles. The fourth column gives the 
    extent of the given
    region along the $y$ axis, and the last column shows the width of the ``guard region'', composed 
    of cells with the full
    number of macro-particles, at both ends of the $x$ axis of the main region, where applicable.}
    \br
    region      & number of   & number of   & transverse        & guard\\
    & electrons   & ions        & extent [\si{\um}] & width [\si{\um}]\\
    \mr
    $N_{n_e}$       & $n_e / n_c$ & $N_{n_e} / Z$       & $(\0 0, \0 5)$      & --    \\
    $N_{n_e/2}$   & $N_{n_e} / 2$   & $N_{n_e/2} / Z$   & $(\0 5, 10)$     & 0.3   \\
    $N_{n_e/10}$  & $N_{n_e} / 10$  & $N_{n_e/10} / Z$  & $(10, 15)$    & 0.2   \\
    $N_{10}$    & $10$        & $2$             & $(15, 20)$    & 0.1   \\
    \br
\endTable

The number of macro-particles varied along the y axis to ensure adequate resolution with $N_{n_e} = n_e/n_c$ electron 
macro-particles per cell in the middle of the simulated target, and save computational time at its far end where the 
background plasma dynamics is less violent. This was achieved by dividing the target into regions, 
schematically depicted in \fref{tgt-structure}, with reduced number 
of particles per cell $N_{\mathrm{ppc}}$ compared to the base value of $N_{n_e} = n_e/n_c$. To maintain the same 
initial electron density $n_e$, these particles have been given an appropriately higher computational weight. Regions 
with lower $N_{\mathrm{ppc}}$, summarized in
\tref{tgt-structure-tbl}, were guarded with a thin layer of cells containing the base number of particles 
$N_{n_e}$ so that the simulated plasma expansion into the vacuum could represent densities lower than those 
represented by the higher weight particles. The target subdivision is the same as in 
\cite{VyskocilSimulationsbremsstrahlungemission2018} apart form the transverse extent of the target which is 
only $y \in (-20, 20)\;\si{\um}$ in this paper. Unless explicitly stated otherwise, we model the 
target as an idealized flat surface foil with no presence of pre-plasma.

\section{Results}
\label{results}

\subsection{Photon spectra}

\begin{figure}[h]
    \figurespacing\includegraphics[width=\fullfigure]{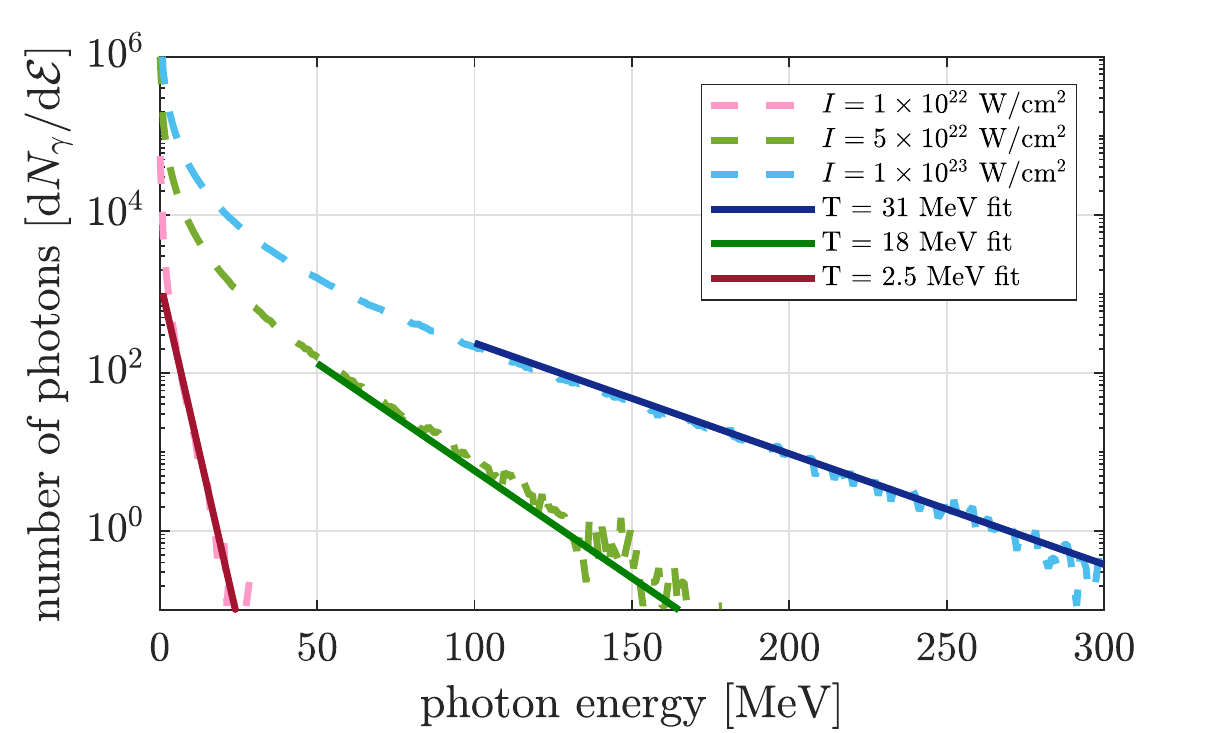}
    \caption{\label{fig-ics-spectrum-ch-int}Spectra of photons radiated via the inverse Compton scattering process from 
        a $d=\SI{2}{\um}$ thick CH foil at three different driving pulse intensities. The tail of each spectrum 
        curve is overlaid with an exponential temperature fit.}
\end{figure}

\Fref{fig-ics-spectrum-ch-int} shows the spectra of all photons generated during the simulation via the inverse Compton
scattering process. The EPOCH algorithm is set up so that the minimum energy of an emitted photon is $\Egamma >
\SI{100}{keV}$, though we limit the analysis to photons with $\Egamma > \SI{1}{MeV}$. In this case, 
there exists a threshold laser pulse intensity $I \simeq \SI{3e21}{W/cm^2}$, corresponding
to $a_0 \simeq 50$ potential, below which no ICS-produced photons are seen in the simulation. The tail of 
this
distribution can be approximated by an exponential temperature fit $N_\gamma \approx \exp\left(-\Egamma/\kB
\Tgamma\right)$, included in the figure. 

\begin{figure}[h]
    \figurespacing\includegraphics[width=\fullfigure]{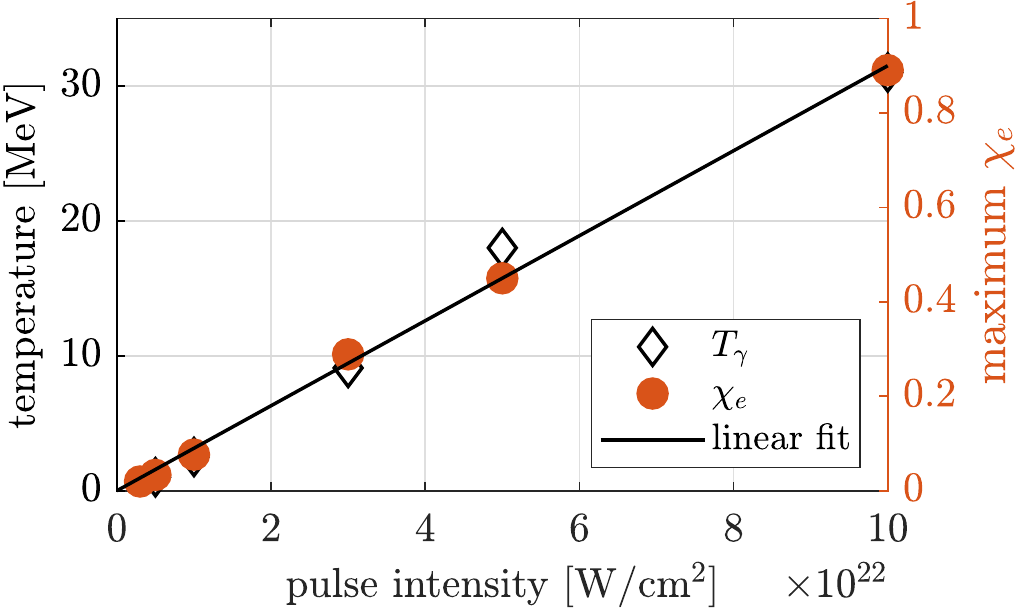}
    \caption{\label{fig-ics-temp-vs-intensity}Effective temperature $\Tgamma$ of the emitted inverse Compton scattering 
    photons, and the maximum emission parameter $\chie$ attained by hot electrons from a $d=\SI{2}{\um}$ CH foil 
    interacting with laser pulses of different intensities.}
\end{figure}

Unlike the bremsstrahlung case \cite{VyskocilSimulationsbremsstrahlungemission2018}, where the effective 
photon temperature is linear in the
potential $\Tgamma^\textup{BS} \sim a_0 \sim \sqrt{I}$, the 
temperature of the simulated photons emitted by the ICS process, shown in 
\fref{fig-ics-temp-vs-intensity}, reveals a scaling linear in the intensity
\begin{equation}
\Tgamma \sim a_0^2 \sim I.
\end{equation}

The $\chie$ parameter
governing the emission process depends on both the velocity of the electron and the strength of the external field $E$ 
to which it is subjected in a given moment, 
\begin{equation}\label{chie-i} 
\chie \sim \gamma E \sim I.
\end{equation}
This follows from the observation that as the ponderomotive scaling equation \eref{wilks-scaling} holds, and the 
$\gamma$
factor attained by the hot electron population $\gamma \sim a_0 \sim \sqrt{I}$, the emission parameter ought to be
proportional to both the gamma factor and the strength of the electric field $E\sim\sqrt{I}$, thus being linearly
dependent on the intensity as shown in \fref{fig-ics-temp-vs-intensity}. Though since the electron temperature is $\Te 
\sim \sqrt{I}$, and the photon temperature must be $\Tgamma < \Te$, there has to be a turning point where 
the 
raise in $\Tgamma$ slows down at some higher intensity, and the scaling $\Tgamma \sim I$ ceases 
to be 
valid.

\begin{figure}[h]
    \centering
    \includegraphics[width=\fullfigurepng]{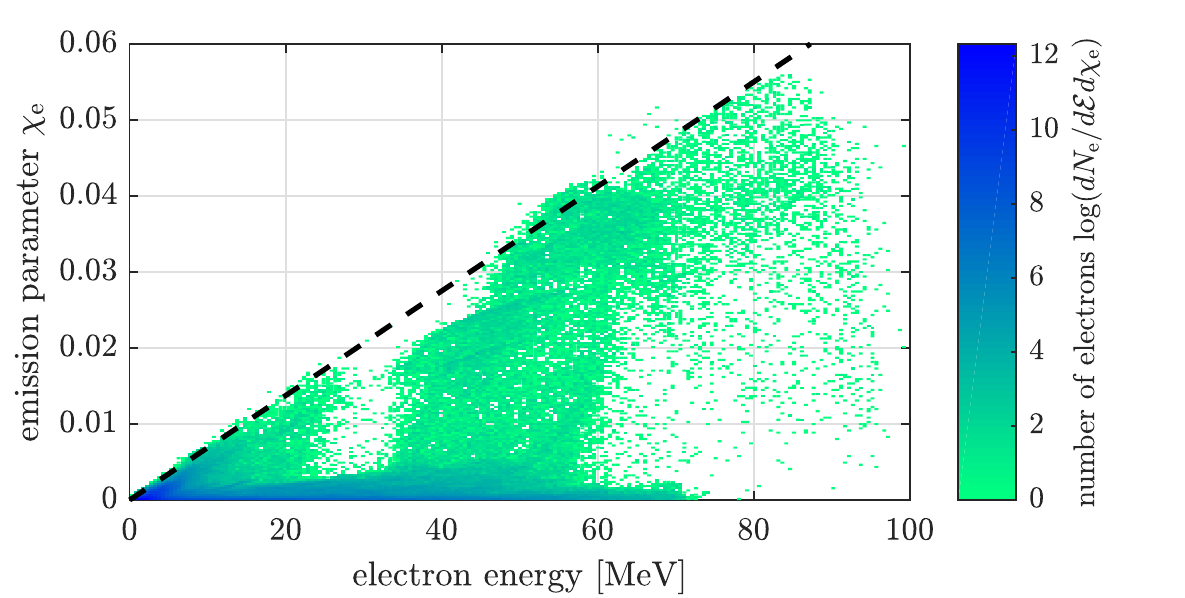}
    \caption{\label{fig-gamma-chi-sim}Emission parameter $\chie$ plotted against the electron $\gamma$ 
    factor from a snapshot taken at the peak emission time of a half-cycle around $t=\SI{110}{fs}$ from a 
    simulation of a $d=\SI{2}{\um}$ CH foil interacting with a $I=\SI{e22}{W/cm^2}$ laser pulse (cf. 
    \fref{snapshots} with a snapshot of the same set of electrons). The colour 
    indicates the number of electron macro-particles of a particular energy and emission parameter in log 
    scale.}
\end{figure}

An estimate for the most common energy of the resulting radiation has been proposed in the monochromatic 
approximation, giving $\hbar\omega_\gamma \simeq 0.44\chie\gamma \me c^2$ 
\cite{BellPossibilityProlificPair2008, KirkPairproductioncounterpropagating2009, 
    RidgersDenseElectronPositronPlasmas2012}.
This expression though describes the 
maximum of the photon distribution while the effective photon temperature $\Tgamma$ comes form a fit of 
the tail of a distribution which covers photons emitted by all of the electrons over the course of the 
simulation, therefore this expression cannot not predict the temperature of the photons based on that of the 
electrons in our situation. As the immediate value of $\chie$ depends on the exact trajectory of the 
electron, a simple connection between the temperature $\Te$ of the accelerated electron bunches and the 
temperature of the resulting radiation $\Tgamma$ cannot be made in the complex case of the laser-solid 
interaction where the bunch is of a finite size and, consequently, the different electrons interact with the 
field in a different phase. This is evident from the snapshot in \fref{fig-gamma-chi-sim}, obtained from 
detailed studies of electron trajectories presented later in \sref{eln-dynamics}, which shows the relation 
between the $\gamma$ factor and the $\chie$ parameter of the simulated electrons. We observe that there are 
many hot electrons which have the same $\gamma$ factor but span a broad range of attained $\chie$. 
Therefore, the immediate electron temperature $\Te$ does not readily reveal the radiation temperature 
$\Tgamma$, though averaging over many samples during the course of the whole interaction where both the 
$\gamma$ factor and the field strength vary with each laser cycle would ultimately lead to a 
Maxwell-Boltzmann like distribution. Though we see that the maximum $\chie$ is linear in electron energy, 
the 
linear scaling $\Tgamma \sim a_0^2 \sim I$, which turns out quite clearly in 
\fref{fig-ics-temp-vs-intensity}, should be treated as an empirical observation.

\subsection{Standing wave model}

The inverse Compton scattering process involves an electron moving in the field of the laser pulse in front of the
target. To obtain more insight into the physical mechanisms governing the emission, we will make use of the simulation 
data with high temporal resolution with the help of a simplified theoretical model derived to 
describe the electron motion based on the standing wave approximation, which will be solved numerically.

As the electromagnetic wave of the linearly polarized laser pulse impinges on the highly overdense flat plasma slab at 
$x = 0$, most of it is reflected back and interferes with the incoming part of the pulse forming a standing 
wave in 
front of the target. The more equal the incident and reflected pulses, the more pronounced the standing wave 
pattern.
In the case of a very short pulse, where the field intensity of the 
envelope changes rapidly with each oscillation, this pattern would be most prominent around the peak of the 
laser-target interaction where the intensity profile of the incoming and the reflected waves are approximately equal. 
The electric and magnetic field of the standing wave formed in front of the target in the case of normal incidence can 
be approximated by a plane wave near the interaction centre, and characterized by:
\begin{equation}\label{sw-eb}
\eqalign{
    \Ey = E_0 \sin(\omega t) \sin(kx), \cr
    \Bz = B_0 \cos(\omega t) \cos(kx),}
\end{equation}
where $B_0 = E_0/c$. At the target's surface, the $\Ey$ field then has a node, while the $\Bz$ field then has an 
anti-node. The maximum amplitude of the standing wave field is twice as large as that of 
the incident pulse due to the constructive interference of its incoming and outgoing parts.

In order to characterize the inverse Compton scattering radiation of an electron injected from the plasma surface into 
the standing wave, we expand equation \eref{chi-e} assuming $\ve{B} = (0, 0, \Bz)$ and $\ve{E} = (0, \Ey, 0)$. 
For high energy electrons with momenta $p \gg \me c$, we can make the approximation $\gamma^2 \simeq 
\left(p/\me c\right)^2$. Furthermore, as there are no forces acting on the electron along the $z$ axis, $\pz = 0$, we 
can take $\gamma^2 \simeq (\px/\me c)^2 + (\py/\me c)^2$
finally obtaining the simplified approximation
\begin{equation}\label{chie-simple}
\chie \approx \frac{1}{\ES} \left|\frac{\px \Ey}{\me c} - c\gamma\Bz\right|.
\end{equation}

\begin{figure}[h]
    \figurespacing
    (a) \includegraphics[width=\halffigure]{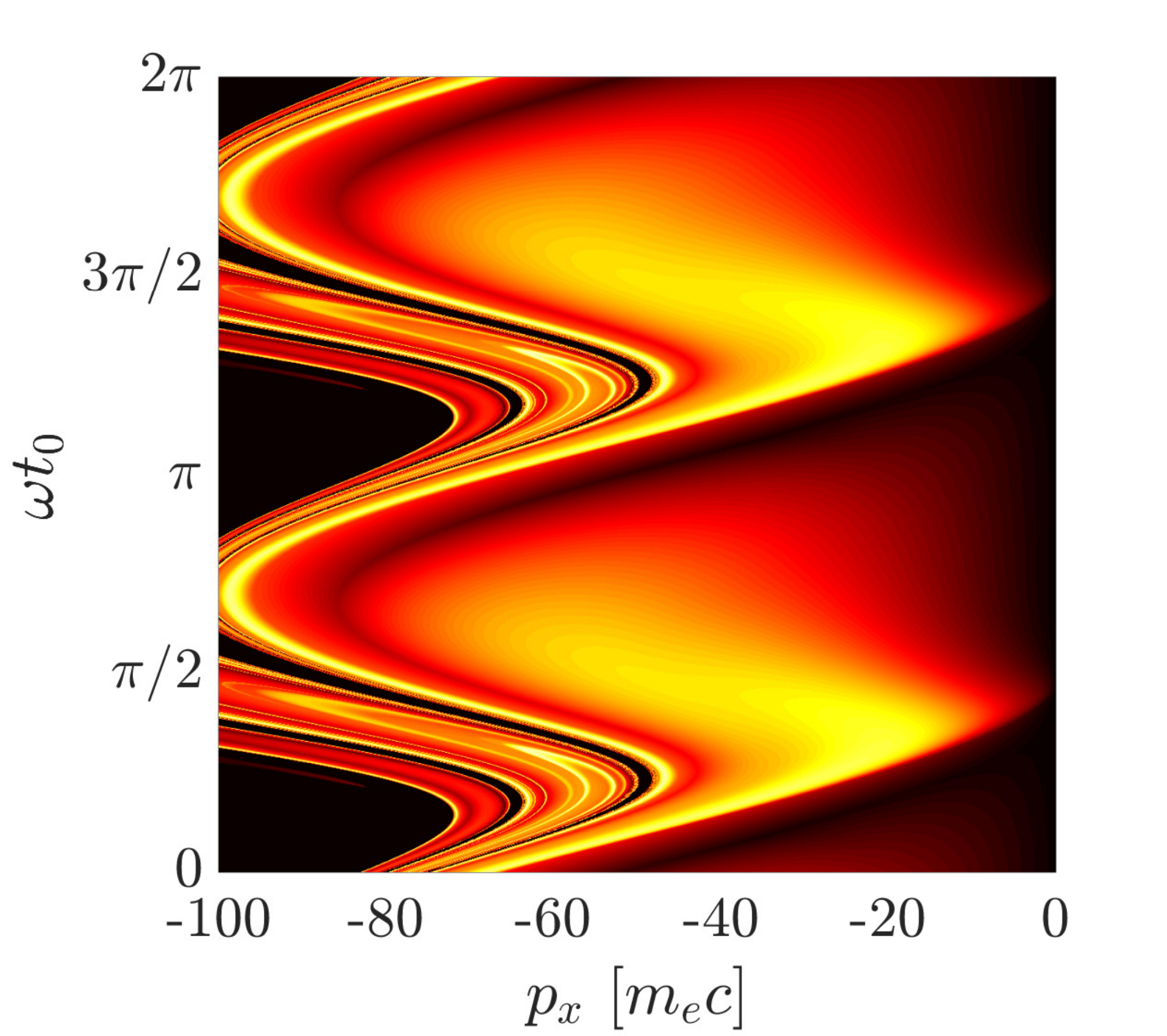}
    (b) \includegraphics[width=\halffigure]{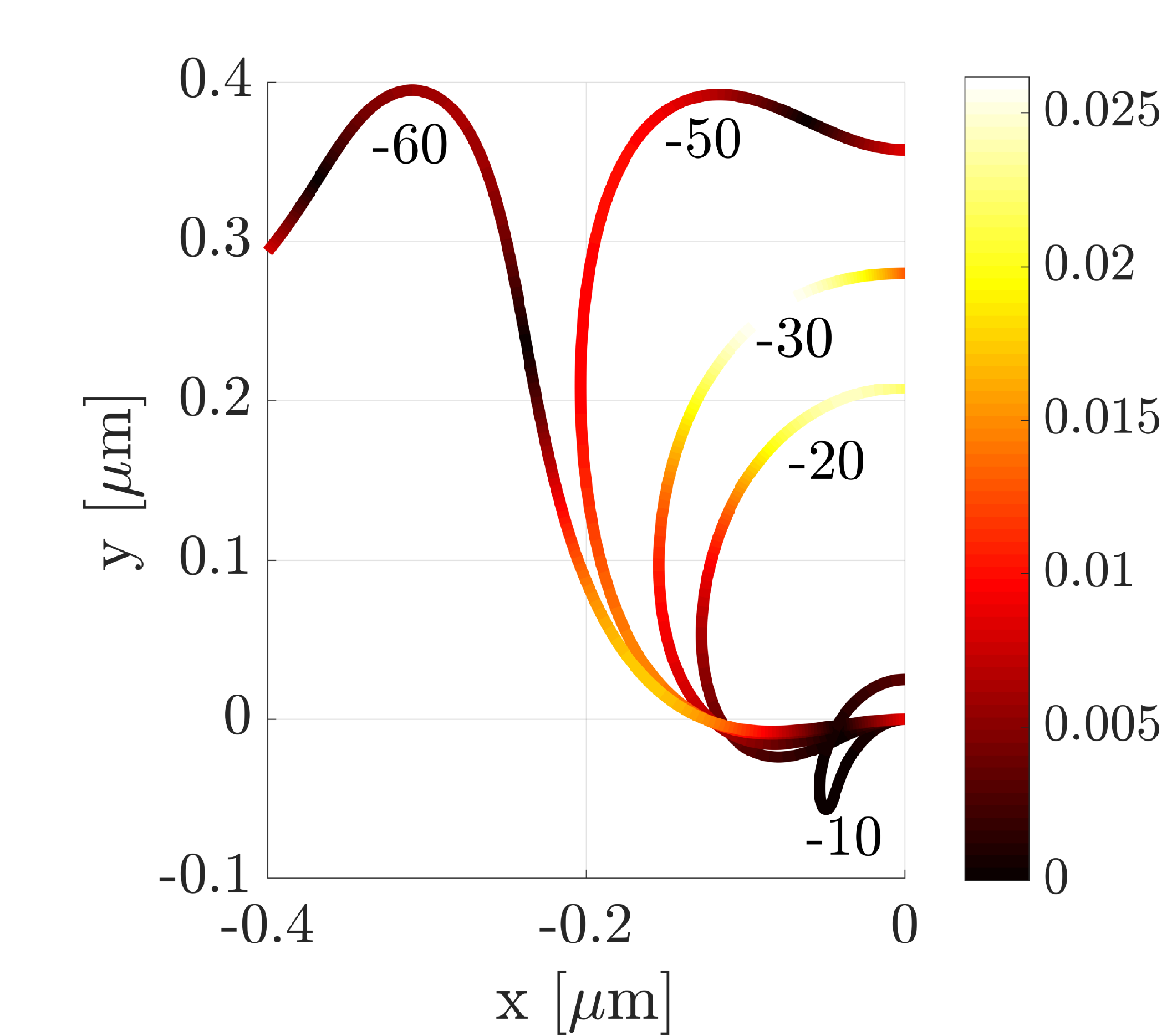}
    
    \caption{\label{model-phase}Numerical prediction of maximum $\chie$ attained by electrons of different initial 
        momenta injected into the standing wave formed by a $I=\SI{e22}{W/cm^2}$ laser pulse at a different phase. In 
        (a), 
        the maximum emission parameter is indicated by the colour. In (b) trajectories of electrons injected with 
        $p_\textup{x,0} = -10, -20, -30, -50$, and $-60\;\me c$ into the $\omega t_0 = 3/\pi$ phase are plotted with 
        the 
        colour indicating the instant $\chie$ of the specific electron at that point.}
\end{figure}

\begin{figure}[h]
    \figurespacing
    \includegraphics[width=\fullfigurepng]{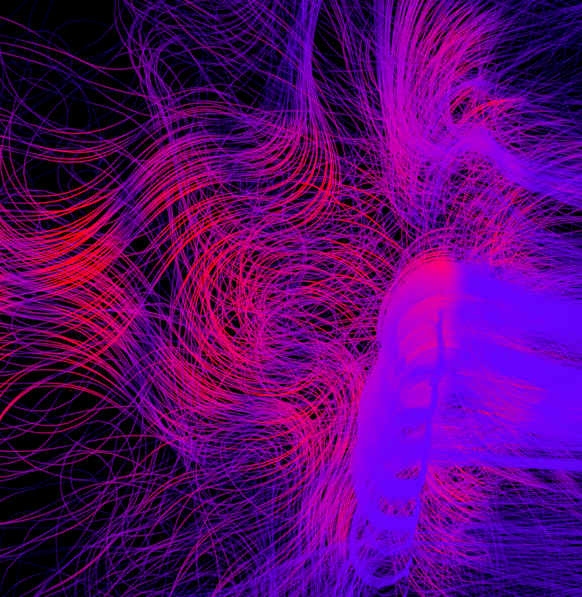}
    
    \caption{\label{mari-tracks}Trajectories of electrons which were accelerated form the front side of a 
        $d=\SI{2}{\um}$ target during one of the early half-cycles of a $I=\SI{e22}{W/cm^2}$ laser pulse. The simulation
        area depicted in the illustration spans approximately $x \in (-1, 0.5)\,\si{\um}$ and $y \in (-1, 1)\,\si{\um}$.
        The colour indicates the instant $\chie$ of the specific electron at that point with the scale going from blue
        (low) to magenta (high). Electrons were selected on the basis of attaining $\chie >0.01$ during one 
        laser pulse 
        half-cycle, then their trajectories were plotted from the beginning of the half-cycle till the end 
        of the 
        simulation. The laser pulse was incoming from the left and injected many electrons into the target
        on an almost half-circle trajectory, seen in the lower right part of the picture. Upon entering the target, the
        electrons are not influenced by any strong fields, and continue in a straight line. After they reflect at the
        back side of the target (far right outside this illustration), some of them, albeit a much lower number,
        re-enter the interaction area with a high initial velocity, and radiate in the backward direction. This
        secondary emission happens at a late time of the interaction, and only those electrons that have had been
        injected in the earliest time arrive soon enough to meet the laser pulse at sufficient intensity to emit any
        significant amount of radiation -- cf.\ total forward vs.\ backward emission in \fref{fig-ics-ang-ch-int}(a).}
        \end{figure}

Equations \eref{sw-eb} and \eref{chie-simple} can be solved numerically, coupled with the relativistic equation of 
motion of the electron. \Fref{model-phase} shows the predicted maximum $\chie$ attained by electrons of different 
initial momenta injected into the standing wave at different phase which radiate in the space in front of 
the target in 
the positive $x$ direction. Around $p_\textup{x,0} = -20\me c$ at a phase below $\omega t_0 = \pi / 2$, 
there is a 
region of stability with respect to these two parameters. Electrons injected with a much lower initial 
momentum do not 
radiate at all, while those with a much higher one will never return into the target, and will radiate in 
the backward 
direction. Such a high momentum injection cannot be achieved by the interaction of the laser pulse with the 
front side 
electrons, and does not appear in the full PIC simulations. However, similar trajectories, depicted in 
\fref{mari-tracks}, can occur when recirculating electrons return form the back side of the target, and 
enter the area 
in front of the target while the pulse has a different phase than it would have had in case of direct 
injection from 
the front side. This kind of backward emission can be seen in very thin $d \le \SI{2}{\um}$ foil in the late 
time of 
the interaction, being caused by the electrons which were injected early, and had enough time to do a 
subsequent full 
revolution in the target. Since the electron bunch spreads out in the transverse direction during the 
recirculation 
process \cite{VyskocilSimulationsbremsstrahlungemission2018}, the 
returning electrons can be seen as essentially sampling arbitrary pulse phases in the $(\omega t_0, \px)$ 
phase-space.

\begin{figure}[h]
    \figurespacing
    (a) \includegraphics[width=\halffigure]{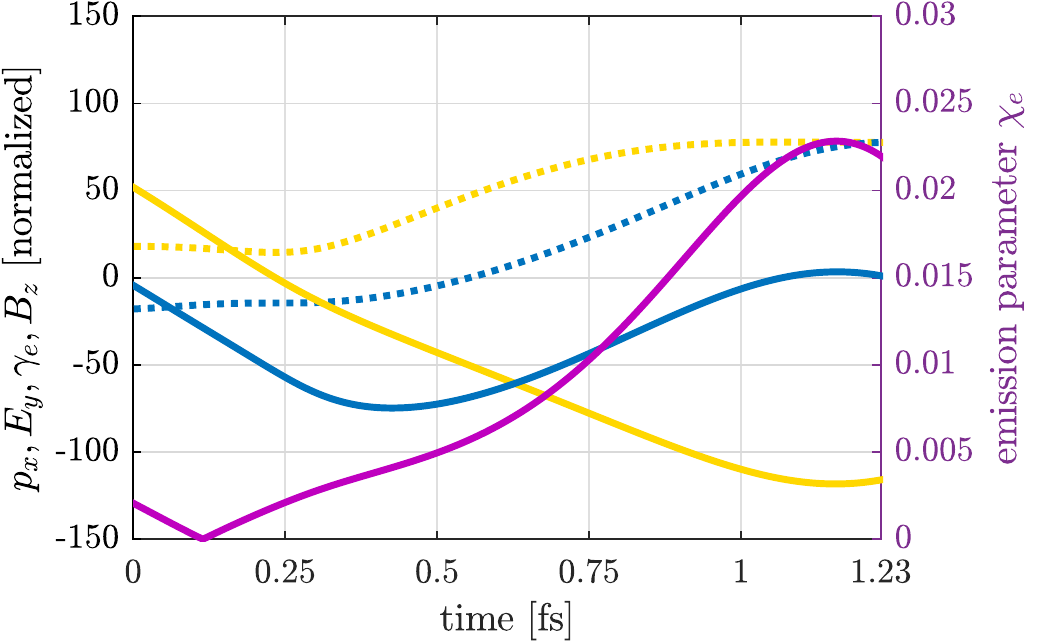}
    (b) \includegraphics[width=\halffigure]{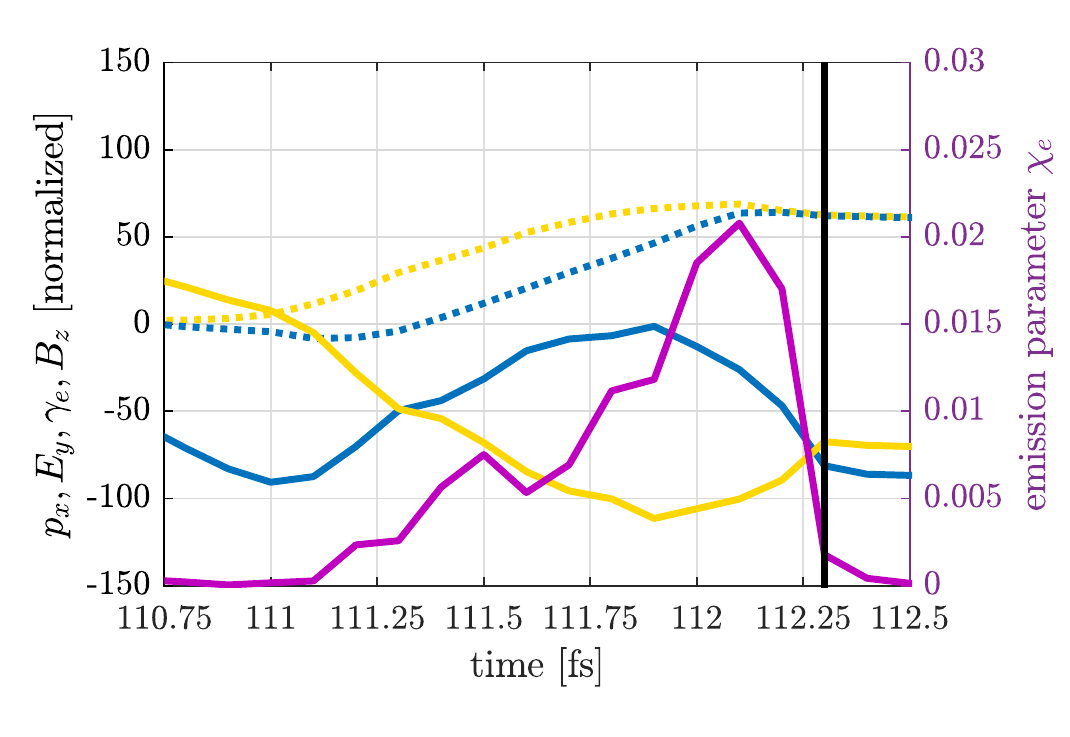}

    \begin{center}
        \includegraphics[width=\halffigure]{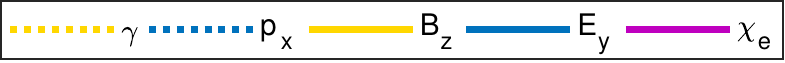}
    \end{center}
    
    \caption{\label{model-time}Time evolution of the momentum $\px$ of an electron, its relativistic $\gamma$ 
        factor, the $\Bz$ and $\Ey$ fields along its trajectory, and the emission parameter $\chie$ (a) from the 
        simplified theoretical model with the initial momentum corresponding to $\E_\textup{e} = \SI{9}{MeV}$ in a 
        standing wave with a peak intensity $I_\textup{SW} = \SI{2e22}{W/cm^2}$ formed by the reflection of a 
        $I=\SI{e22}{W/cm^2}$ pulse, and (b) from the Particle-in-Cell simulation of a $I = \SI{e22}{W/cm^2}$ laser 
        pulse interacting with a $d=\SI{2}{\um}$ CH foil. The momentum $\px$ is in the normalized units of $[\si{\me 
        c}]$, the electric field $\Ey$ in $[\si{\me \omega c/e}]$, and the magnetic field
        $\Bz$ in $[\si{\me \omega/e}]$. Note that the calculation of the theoretical model stops when the electron
        re-enters the target, while no such limit exists in the PIC simulation, therefore we track the electron after 
        being re-injected to show that $\chie$ indeed drops to zero.}
\end{figure}

For a sample numerical solution, we calculated the time evolution of the model for the initial momentum of 
$p_\textup{x,0} = -18\me c$, which corresponds to the energy $\E_\textup{e} = \SI{9}{MeV}$, injected into the $\pi/3$ 
phase of a standing wave with peak intensity $I_\textup{SW} = \SI{2e22}{W/cm^2}$, which corresponds to the constructive 
interference of the incoming and reflected parts of an $a_0 = 86, I = \SI{e22}{W/cm^2}$ laser pulse. The electron's 
trajectory starts and ends at the surface of a target positioned at $x = 0$. The model tracks the evolution of the 
electric $\Ey$ and magnetic $\Bz$ fields along the trajectory of the simulated electron. Together with the electron's 
$\px$ and its $\gamma$ factor, these constitute the two parts of the simplified equation \eref{chie-simple}. The 
result, shown in \fref{model-time}(a), compares favourably to the actual trajectory of an electron, in 
\fref{model-time}(b), selected form the PIC simulation on the basis of similar injection phase, and the 
initial and final relativistic $\gamma$ 
factor.

\subsection{Electron dynamics}\label{eln-dynamics}

\begin{figure}[h]
    \figurespacing
    \includegraphics[width=\halffigure]{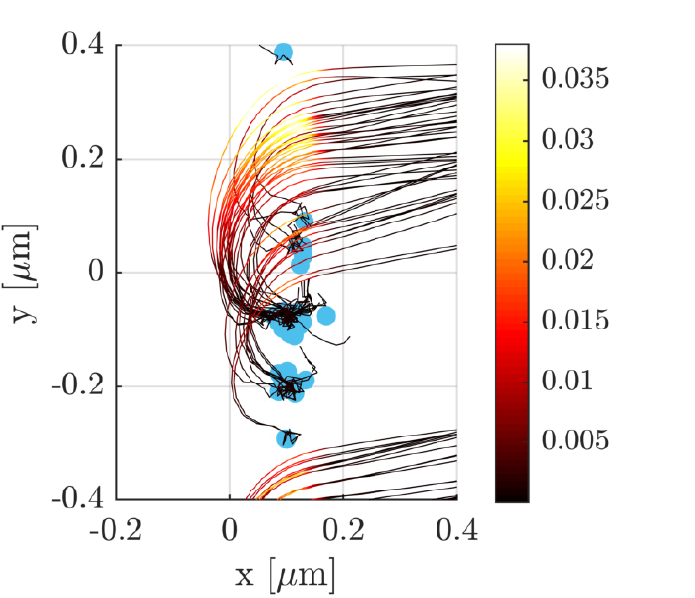}
    
    \caption{\label{snapshots} The blue dots represent positions of the particles from the PIC simulation of a 
        $d=\SI{2}{\um}$ CH foil, right before they are pulled out of the target, overlaid on the trajectory followed by 
        the respective particle during one half-cycle of a $I = \SI{e22}{W/cm^2}$ laser pulse. The 
        trajectory colour 
        shows the value of the 
        emission parameter $\chie$ achieved by the electron along the trajectory.}
\end{figure}

In order to describe the dynamics of the electrons responsible for the gamma ray emission via inverse Compton
scattering, we can compare the results of the numerical solution of equations \eref{sw-eb} and \eref{chie-simple}, seen 
in \fref{model-time}(a), to a simulation snapshot zoomed-in to the centre of the interaction area in \fref{snapshots}. 
It shows the trajectories of a random sample of electrons which achieve a high value of $\chie$ during one half-cycle 
of the driving laser pulse. In the simulation, a total of about 18~000 electron macro-particle reach 
$\chie>0.01$ 
during this particular half-cycle, and over 99\% of them follow trajectories of a similar shape as the one 
produced by 
the aforementioned sample numerical solution. At this stage of the interaction, hole boring by the laser 
pulse has 
pushed the target surface from $x=0$ to $x\approx\SI{150}{nm}$, the phase of the $\Ey$ field is changing, 
and a new 
bunch is about to be accelerated.

First, the electron is pulled out of the target surface, and injected into the standing wave in front of the target 
when the balance between the $\ve{J} \times \ve{B}$ force and the force due to the $\Ex$ field is violated. This stage 
is not covered by the theoretical model, where we instead inject the electron with a specified initial momentum (or a 
range of momenta, as will be described in the following text), and neglect the $\Ex$ field altogether.

After being injected, the electron is accelerated in the $+y$ direction by the $\Ey$ field, causing a rise in its 
relativistic $\gamma$ factor. Meanwhile, the phase of the $\Bz$ field changes, causing the increase in the originally 
negative momentum $\px$ up to a moment when $\px=0$, and the electron is at the maximum distance $\Delta 
x\approx\SI{180}{nm}$ away from the actual target surface.

Next, the rising $\Bz$ field transforms the transverse momentum $\py$ into the longitudinal $\px$ as the electric field 
$\Ey$ weakens. The relativistic $\gamma$ factor is dominated by the $\px$ component -- in the normalized units of 
\fref{model-time}, $\px\approx\gamma$, and the electron is returning into the target with $\px \gg \py$. 

Maximum $\chie$ parameter is attained right before the re-injection, when $\gamma$ is almost constant as $\Ey$ is 
decreasing with the impending phase change. The $c\gamma\Bz$ is now the dominant term in equation \eref{chie-simple}, 
but due to the still non-negligible $\py$, maximum emission occurs at an angle $\alpha \neq 0$. Right before 
the 
re-injection, the $\Bz$ field starts to decrease, and the electrons, which have lost most of their 
transverse momentum 
continue to propagate inside the target. The process is about to repeat with the forthcoming laser pulse 
half-cycle, 
albeit mirrored with respect to the $x$ axis.

\subsection{Emission angle}

\begin{figure}[h]
    \figurespacing
    \includegraphics[width=\fullfigure]{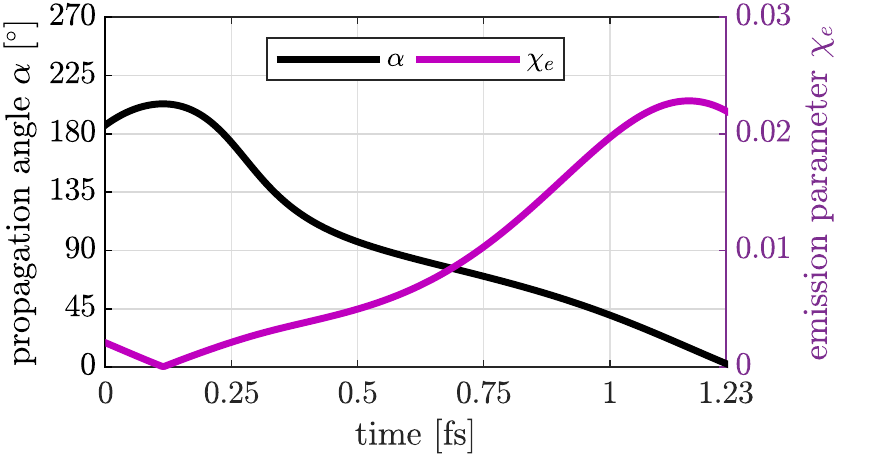}
    \caption{\label{model-time-ang}Time evolution of the emission parameter $\chie$, and the propagation angle 
        $\alpha$, measured for the $x$ axis, of the emitting electron in the simplified
        theoretical model with the initial momentum corresponding to $\E_\textup{e} = \SI{9}{MeV}$ injected 
        into the 
        $\omega t_0 = \pi/3$ phase of a standing wave with a peak intensity $I_\textup{SW} = 
        \SI{2e22}{W/cm^2}$, formed 
        by the reflection of a $I=\SI{e22}{W/cm^2}$ pulse.}
\end{figure}

\begin{figure}[h]
    \centering
    \includegraphics[width=\fullfigure]{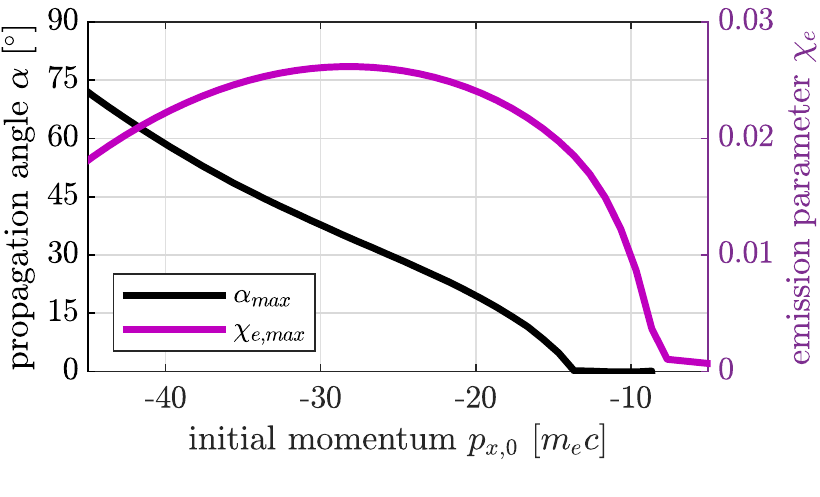}
    \caption{\label{model-a86}Theoretical model of the maximum emission parameter $\chie^\textup{max}$, and the 
        corresponding emission angle $\alpha_\textup{max}$ for different initial energies of electrons 
        injected into 
        the $\omega t_0 = \pi/3$ phase of a standing wave with a peak intensity $I_\textup{SW} = 
        \SI{2e22}{W/cm^2}$ 
        formed by the reflection of a $I=\SI{e22}{W/cm^2}$ pulse.}
\end{figure}

As we have seen that the maximum emission occurs when the electron is propagating at an angle, we shall now discuss 
some features of the angular distribution of the emitted photons seen in the theoretical model. \Fref{model-time-ang} 
shows that the theoretical model predicts an angle $\alpha_\textup{max}$, measured from the $x$ axis, where the 
emission parameter $\chie$ has a maximum for an electron with a given initial momentum. To see how the angle of maximum 
emission changes in case when a spectrum of electrons would be injected, we first calculate the model values for a 
range of initial electron momenta. For each energy, we find the time $t_\textup{max}$ when the emission parameter has a 
maximum $\chie^\textup{max} = \chie(t_\textup{max})$, $\textup{d}\chie/\textup{d}t|_{t_\textup{max}} = 0$, and the 
angle $\alpha_\textup{max} = \alpha(t_\textup{max})$ at which the maximum emission occurs for the given electron 
energy. \Fref{model-a86} shows that there is an optimal initial electron energy $\E_\textup{e}^\textup{opt}$ which 
leads to the highest value of the emission parameter at a given laser pulse intensity. Electrons around this optimum 
are responsible for the majority of the gamma radiation, while those which are too far away, be they slower or faster, 
would emit considerably less.

\begin{figure}[h]
    \figurespacing
    \includegraphics[width=\fullfigure]{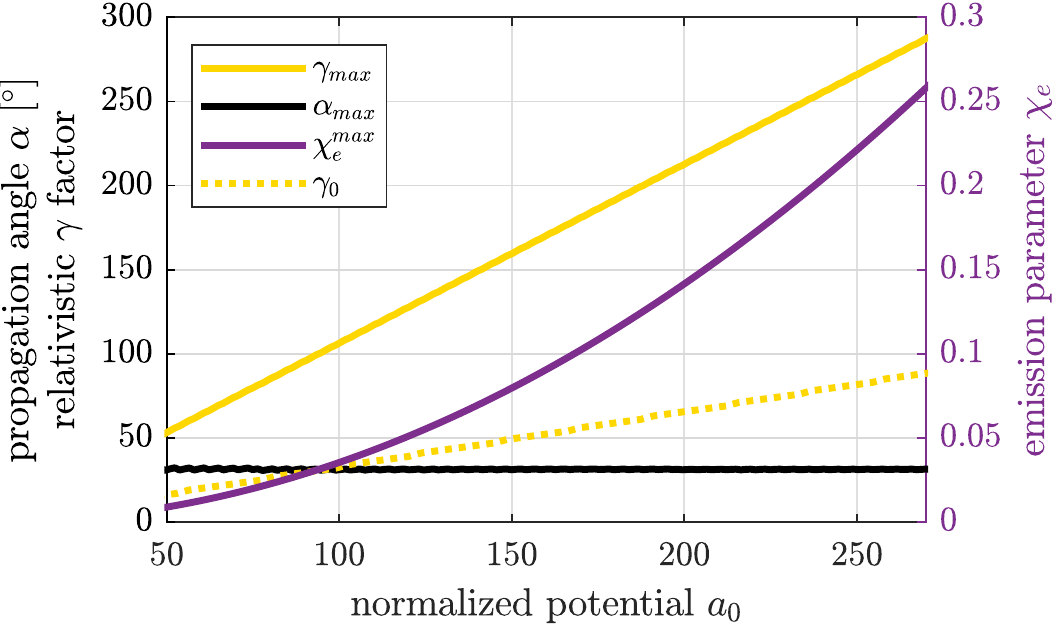}
    \caption{\label{chie-intensity}Theoretical model of the maximum emission parameter $\chie^\textup{max}$ reached by 
    any electron for a given laser pulse potential $a_0 = \sqrt{I}$ (where the standing wave maximum 
    intensity is 
        $I_\textup{SW} = 2I$) with the angle $\alpha_\textup{max}$ at which the emission occurs, the relativistic 
        factor $\gamma_\textup{max}$ attained by the electron at the point of maximum emission, and the 
        relativistic 
        factor $\gamma_\textup{0}$ with which has the electron been injected into the standing wave.}
\end{figure}

Then, we perform a parameter scan over laser pulse intensities, finding the optimal initial electron energy 
$p_\textup{x,0}^\textup{opt}(a_0)$, the emission parameter $\chie^\textup{max}$, the emission angle 
$\alpha_\textup{max}$, and the 
maximum $\gamma$ factor attained by the emitting electron. \Fref{chie-intensity} shows that the maximum $\gamma$ factor 
is linear in $a_0$, thus the maximum emission parameter increases with $\chie \sim a_0^2$. The angle at the 
moment when 
the emission parameter reaches its maximum does not depend on the intensity, and is $\alpha_\textup{max} = 
\ang{30}$. 
If our assumptions hold, one can expect this to be the direction of maximum emission of the ICS gamma rays 
in the 
simulations.

\subsection{Angular distribution}

In the PIC simulations, the angular distribution of photons emitted via the inverse Compton scattering process in the 
interaction with a $I=\SI{e22}{W/cm^2}$ pulse has a distinct structure with two lobes centred around $\vartheta \simeq 
\ang{30}$ and $\vartheta \simeq \ang{330}$. This result is consistent both with 
previously published simulations \cite{NakamuraHighPowergRayFlash2012, JiEnergypartitiongray2014, 
    ChangUltraintenselaserabsorption2017}, and the theoretical model presented in \sref{eln-dynamics}.

\begin{figure}[h]
    \figurespacing
    (a) \includegraphics[width=\halffigure]{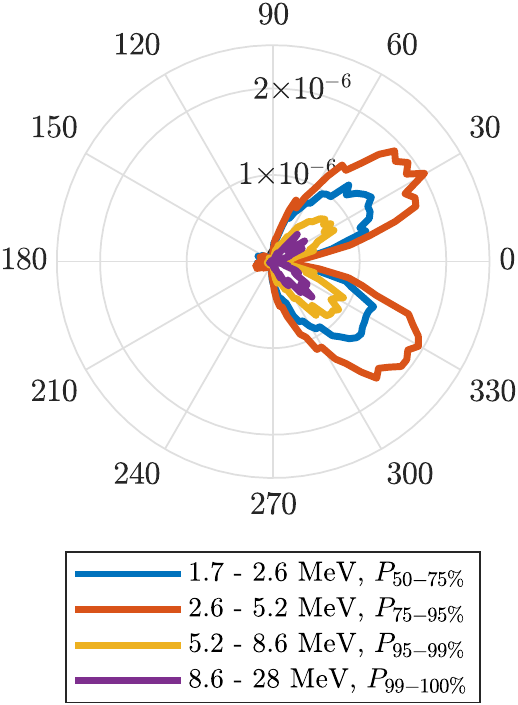}
    (b) \includegraphics[width=\halffigure]{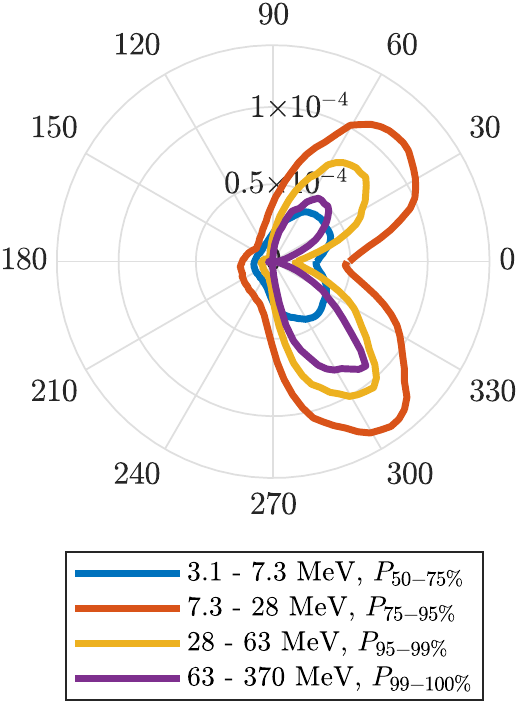}
    
    \caption{\label{fig-ics-ang-ch-int}Angular distribution of photons emitted via the inverse Compton scattering 
    process from CH foils with $d=\SI{2}{\um}$ at different driving pulse intensities $I = \SI{e22}{W/cm^2}$ (a) 
    and $I = \SI{e23}{W/cm^2}$ (b). The different curves 
    represent the sum of the energies of all photons in respective energy span in the units of conversion 
    efficiency of 
    the total laser pulse energy into gamma rays in that energy span in given direction per \ang{1} shown on the 
    radial axis. The selection of energy bands in the figures here is not fixed, but differs between simulations 
    to represent exclusive percentile ranges, indicated in the figure legend, to highlight the similarities of the 
    structure of the spectra which, for different intensities, appear at different absolute energy values.}
\end{figure}

In the case of very thin foils $d < 2c\tau$, recirculating electrons have enough time to make a full
revolution and return to the front side of the target while the interaction with the laser pulse is still ongoing. This
then leads to an appearance of backward radiation, which is suppressed for thicker foils. The 
$d=\SI{2}{\um}$ 
target therefore shows a small amount of backward radiation caused by lower energy electrons injected 
into the 
target early by the rising part of the pulse, as seen in \fref{fig-ics-ang-ch-int}(a) Otherwise, since the 
ICS
photons are only emitted from the area in front of an opaque foil target, the angular structure of the resulting
radiation does not depend on the target thickness. However, at high intensities, it depends on the target material.

\begin{figure}[h]
    \figurespacing
    \includegraphics[width=\fullfigure]{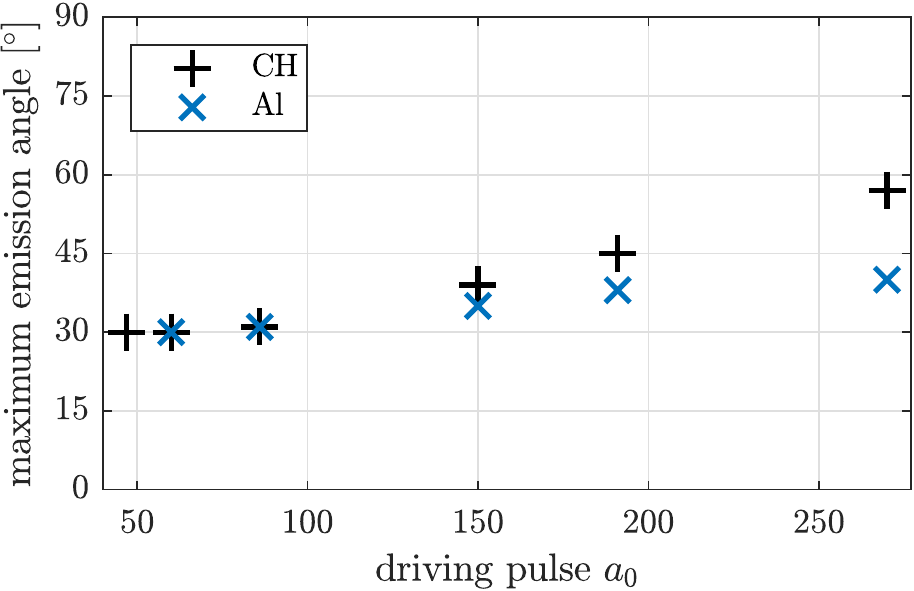}
    \caption{\label{fig-ics-angle-vs-intensity}The most prominent direction of propagation of the high energy photons 
        described as the mode of the angular distribution of all photons with energies above the 50th percentile  for 
        $d=\SI{2}{\um}$ foils from CH and aluminium for driving pulses of different intensities. For $a_0 = 50$, the 
        50th percentile corresponds to $\Egamma \sim \SI{1}{MeV}$, while for $a_0 = 270$, the limit is $\Egamma \sim 
        \SI{3}{MeV}$. }
\end{figure}

\Fref{fig-ics-angle-vs-intensity} shows that most common direction in which the high energy photons radiate, which is 
expressed as the mode of the angular distribution of all photons in the 50th energy percentile, corresponds to the 
theoretical model with $\vartheta \simeq \ang{30}$ up to $I = \SI{1e22}{W/cm^2}$. Then, the angle starts to increase, 
growing faster in the lighter CH foil. This suggests a connection to the hole boring process which is faster at both 
the high intensities and low-Z targets.

The geometry of the front side is defined by the hole boring process 
\cite{RobinsonRelativisticallycorrectholeboring2009} since we do not observe any significant decoupling
\cite{BradyLaserAbsorptionRelativistically2012} of the ion and electron fronts. As the plasma is being 
pushed forward, the depth of the ion front increases gradually in the transverse direction 
towards the centre forming an angled side-wing which stretches from near the focus centre at $y=0$, where the hole 
reaches the maximum depth, to the region with much lower pulse intensity several micrometers away form the centre, 
where the original target surface is virtually undisturbed.

As the intensity increases, faster hole boring leads to a larger incidence angle at the sides of the hole, and we 
cannot assume that the
electrons are pulled in front of the target in the direction normal to the polarization of a standing wave. Instead, 
some enter the interaction area at higher angles. While the radiation is
still predominantly forward-going even for the highest intensity $I=\SI{e23}{W/cm^2}$ examined in this paper, with
increasing intensity, the emission angle increases, backward radiation is enhanced, and the shape of the resulting
spectrum, shown in \fref{fig-ics-ang-ch-int}(b), is approaching that of ``transversely oscillating electron synchrotron
emission'' (TOEE) \cite{ChangUltraintenselaserabsorption2017}, which itself, in simulations parametrised on plasma
density, can be seen as an intermediate stage between the emission from a highly overdense
\cite{RidgersDenseElectronPositronPlasmas2012} and a near-critical-density \cite{BradyGammarayemissioncritical2013,
    BradySynchrotronradiationpair2014} target. Detailed exploration of such low density regimes is out of scope of this
paper, nevertheless the highest-intensity case presented here bears some similarity to the TOEE process. 
Furthermore, in 
this 
high-intensity short pulse interaction, carrier envelope phase effect leads to a pronounced asymmetry of the 
emitted 
radiation.

\begin{figure}[h]
    \figurespacing
    \includegraphics[width=\fullfigure]{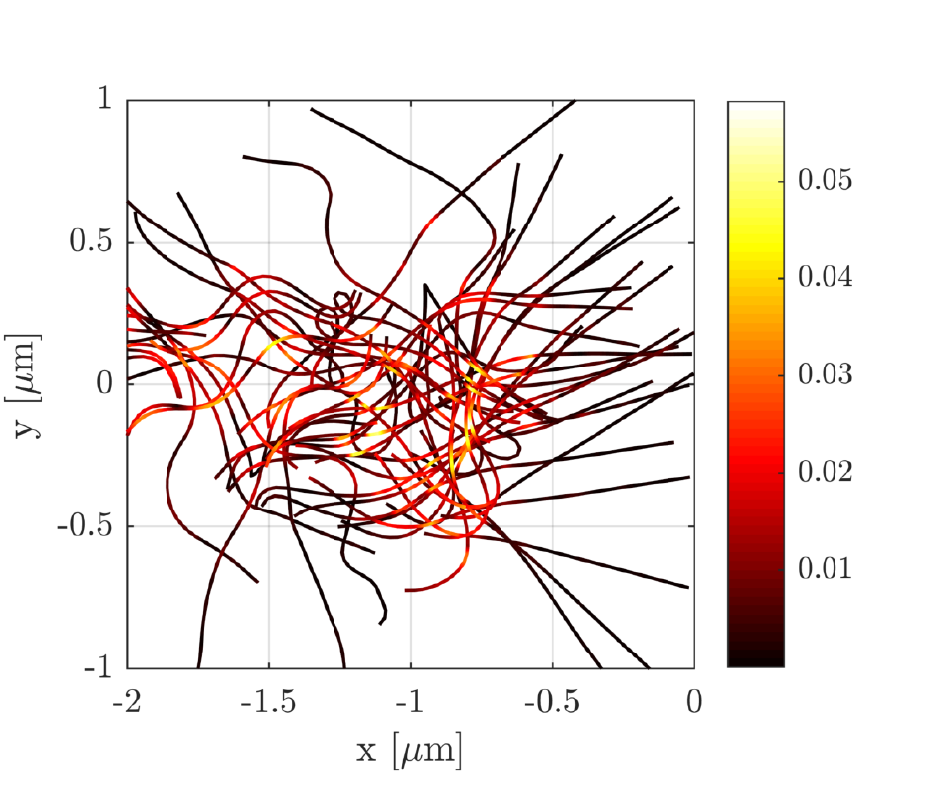}
    
    \caption{\label{cs-tr-pp}Omnidirectional emission in the presence of pre-plasma is due to the 
        chaotic trajectories of electrons such as those seen in this trajectory snapshot taken during one 
        driving 
        pulse half-cycle (cf. \fref{snapshots}). The curve colour represents the immediate value of $\chie$ 
        of a 
        given electron at a given point.}
\end{figure}

\begin{figure}[h]
    \figurespacing
    \includegraphics[width=\halffigure]{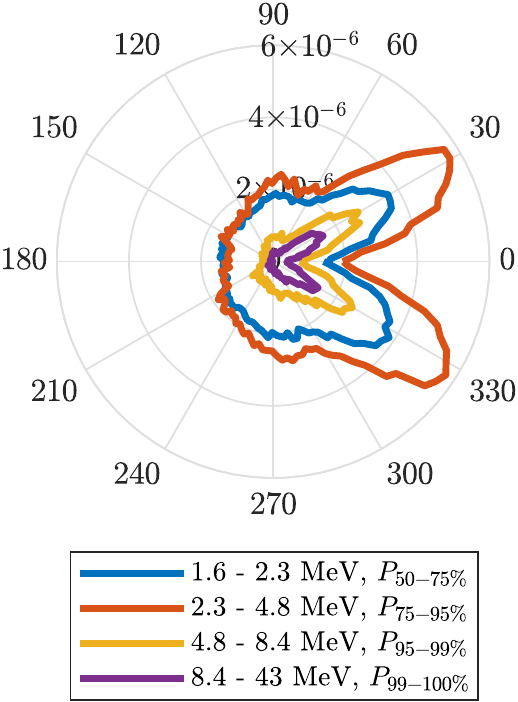}

    \caption{\label{cs-ang-pp}Angular distribution of photons emitted via the inverse Compton scattering 
        process from a $d=\SI{2}{\um}$ CH foil with an exponential pre-plasma profile with the scale length of 
        $l=\SI{1}{\um}$. The different curves represent the sum of the energies of all photons in respective 
        energy span in the units of conversion efficiency of the total laser pulse energy into gamma rays in that 
        energy span in given direction per \ang{1} shown on the radial axis. The omnidirectional, nearly 
        isotropic, 
        emission is due to the chaotic trajectories of electrons such as those seen in the trajectory 
        snapshot in 
        \fref{cs-tr-pp} taken during one driving pulse half-cycle.}
\end{figure}

While the hole boring process influences the gamma ray angular distribution in the case of a solid foil with a flat 
surface, an even more profound effect is revealed in simulations which include pre-plasma, where the 
interaction moves to a 
regime of a laser pulse propagating through underdense plasma. This stage is characterized by side injection from a 
higher density plasma edge formed by electrons pushed away by the ponderomotive force into positively charged channel. 
Energy stored in the space charge field is then released as periodic pulses of backwards propagating electrons which 
are in turn slowed by the radiation reaction force 
\cite{BradyGammarayemissioncritical2013} and emit high energy photons in the backward direction. This process is called 
``reinjected electron synchrotron emission'', or RESE \cite{BradyLaserAbsorptionRelativistically2012}. For an exponential pre-plasma profile with the scale length of $l=\SI{1}{\um}$, the trajectories of the electrons injected from the lower density regions are chaotic, as seen in \fref{cs-tr-pp}, with no readily identifiable typical features. When the laser 
pulse reaches the overdense target, hole boring and reflection occur as in the case without pre-plasma, emitting a 
similar spectrum with the angular distribution featuring the two forward lobes at approximately $\pm\ang{30}$. 
The resulting angular distribution, shown in \fref{cs-ang-pp} is a combination of both processes. Moreover, since the 
electrons are accelerated to higher energies in lower density plasma, the emission is enhanced even in the forward 
direction, where it retains the original structure.

\subsection{Conversion efficiency}\label{ics-ceff}

\begin{figure}[h]
    \figurespacing
    \includegraphics[width=\fullfigure]{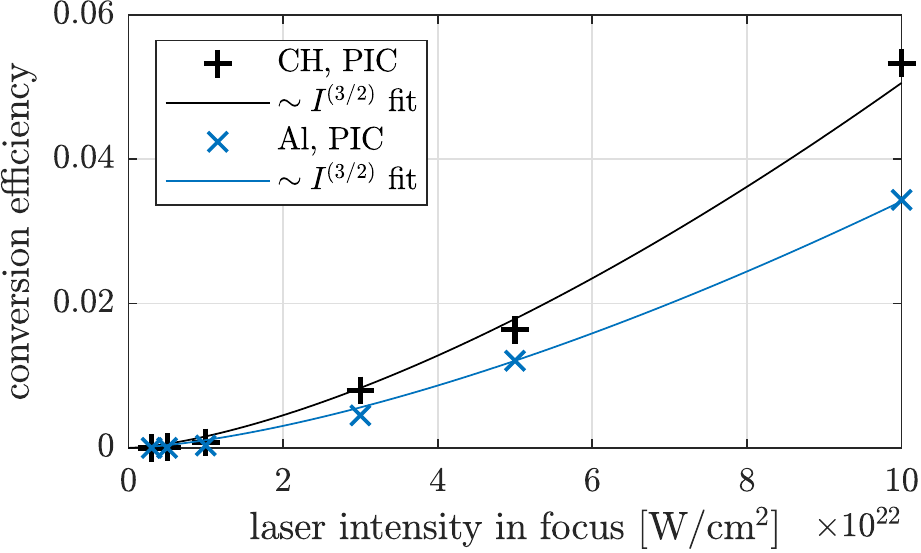}
    \caption{\label{fig-ics-conversion-ch-int}Efficiency of conversion of the total laser pulse energy into all photons 
        emitted via the inverse Compton scattering process from a $d=\SI{2}{\um}$ CH and Al foils as a function of the 
        intensity of the driving laser pulse.}
\end{figure}

\Fref{fig-ics-conversion-ch-int} shows that in our simulations, the total conversion efficiency obeys the scaling 
\begin{equation}\label{eqn-ics-ceff}
\etas{ICS} \sim I^{3/2}
\end{equation}
for both the aluminium and the CH targets. Similar efficiency dependence has been observed in other simulations
\cite{JiEnergypartitiongray2014}. As we have established, in equation \eref{chie-i}, the emission parameter 
scales linearly with the laser pulse intensity, $\chie \sim I$. According to equations \eref{rad-int-low} and 
\eref{rad-int-high}, 
the gamma radiation intensity scales as $I_\textup{rad} \sim \chie^\zeta$ with the power $\zeta = 2$ for 
$\chie \gg 1$, and $\zeta = 2/3$ for $\chie \ll 1$. Our simulations reach up to $\chie \approx 1$, a region 
where neither of the proposed limits are valid. On the one hand, should we lower the intensity to attain 
$\chie \ll 1$, no ICS emission would be seen at all. On the other, with much higher intensities where $\chie 
\gg 1$ would be attained, we can no longer speak about an interaction with an opaque over-critical target 
because of the onset of relativistic transparency. Since we have $\chie \sim I$, equation 
\eref{eqn-ics-ceff} suggests that the region in question could be reasonably described by an intermediate 
empirical value of $\zeta = 3/2$. 

\subsection{Comparison to Bremsstrahlung}

In an experiment, the detectors themselves cannot distinguish between the gamma rays emitted due to bremsstrahlung, 
which we explored in a previous paper \cite{VyskocilSimulationsbremsstrahlungemission2018}, and those emitted due to 
the inverse Compton scattering process studied here. Both will be seen at the same time, and the distinction has to be 
based on distilling their unique features from the total spectra.

\begin{figure}[h]
    \figurespacing
    (a) \includegraphics[width=\halffigure]{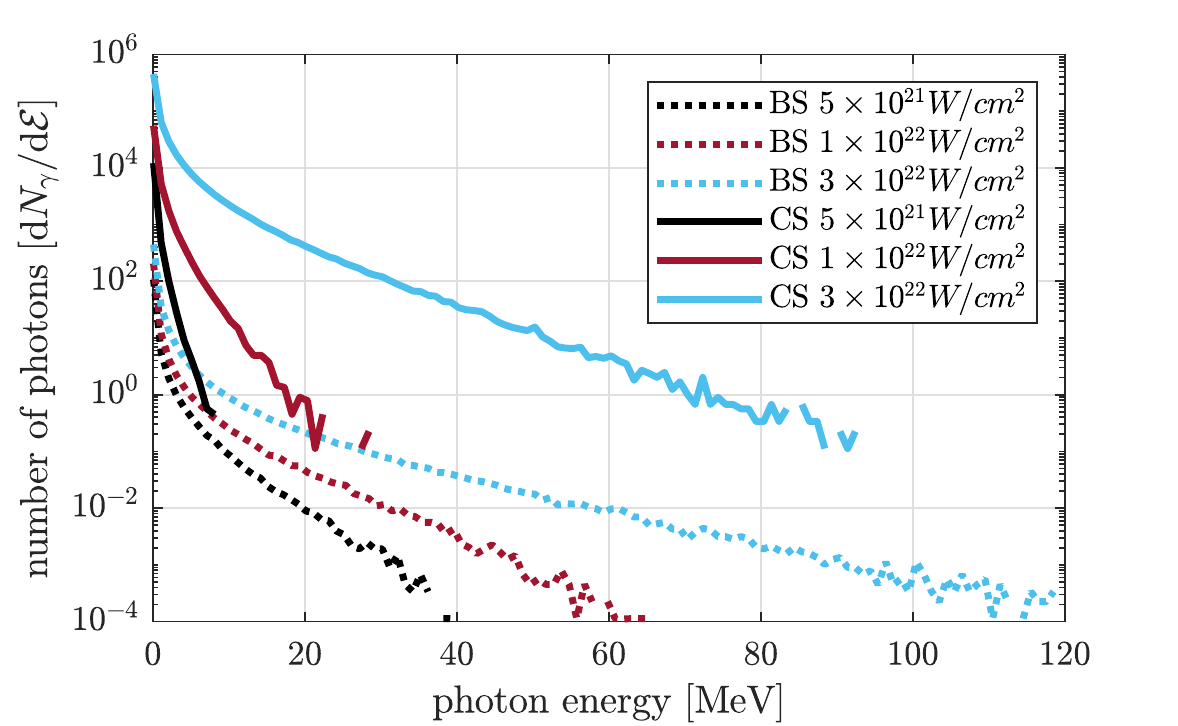}\label{fig-bs-vs-ics-spectrum-int}
    (b) 
    \includegraphics[width=\halffigure]{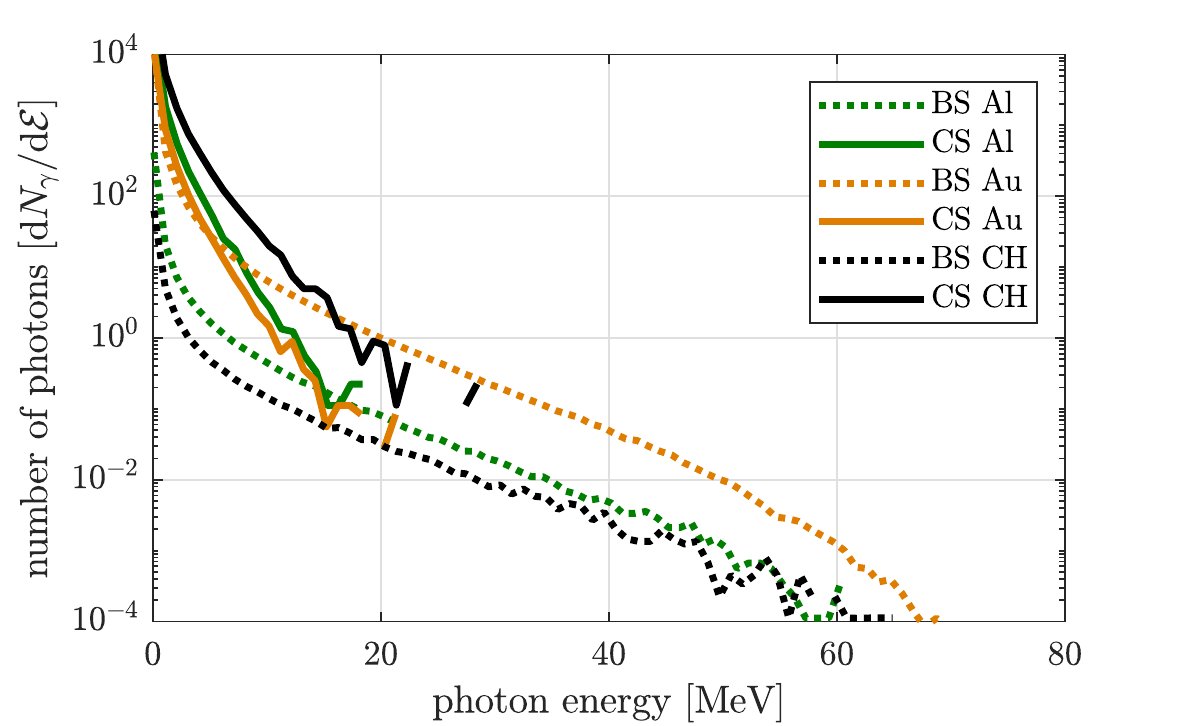}\label{fig-bs-vs-ics-spectrum-mat}
    
    \caption{\label{fig-bs-vs-ics-spectrum}Comparison of bremsstrahlung and inverse Compton scattering spectra in 
        interactions of (a) laser pulses of different intensities with a $d=\SI{2}{\um}$ thick CH foil, and (b) a $I = 
        \SI{e22}{W/cm^2}$ laser pulse with $d=\SI{2}{\um}$ thick foils made of different materials -- 
        C\textsuperscript{6+}H\textsuperscript{+}, Al\textsuperscript{13+}, and Au\textsuperscript{51+}.}
\end{figure}

The first question to be answered is whether the radiation generated by the respective processes would be seen at all. 
In \fref{fig-bs-vs-ics-spectrum}(a), we see that for CH foils, ICS dominates already at the lowest intensity 
$I = 
\SI{5e21}{W/cm^2}$ where it is detectable. Both its temperature and the number of generated photons rise 
quickly with 
the rising intensity, much faster than that of bremsstrahlung. The combination of a thin low-Z target 
irradiated by 
such a high intensity pulse clearly favours ICS. As the bremsstrahlung cross section has a strong dependence 
on the 
atomic number, rising approximately with $Z^2$, using heavier materials should push it to more prominence. 
Actually, as 
seen in \fref{fig-bs-vs-ics-spectrum}(b), the spectrum of bremsstrahlung coming from the Au target dominates 
over that 
of the ICS at the laser pulse intensity of $I=\SI{e22}{W/cm^2}$. A more precise summary of the measured 
values, shown 
in \tref{conversion-material-compare}, reveals that the $d=\SI{2}{\um}$ Au foil is indeed a cross point 
where the total 
conversion efficiencies of the two processes are comparable. Similarly, a comparison can be made between the 
ICS 
emission from the $d=\SI{2}{\um}$ Al foil, and the bremsstrahlung emission from a $d=\SI{5}{\um}$ Al 
foil. Additionally, the effect of lowered absorption and hence a much lower conversion efficiency into the ICS gamma 
rays due to lower electron density can be seen in comparison between the different ionizations of the Au foil.

\begin{table}[ht]
    \centering
    \begin{tabular}{llll@{}}
        \br
        &&		\multicolumn{2}{c}{in [$10^{-6}$] at $t=\SI{170}{fs}$}	    \\
        material    & thickness &  \0\0 $\etas{BS}$ & \0 $\etas{ICS}$ \\
        \mr
        C\textsuperscript{6+}H\textsuperscript{+}     & \SI{2}{\um}    &  \0\0\0 1.6 & \0 690 \\
        & \SI{5}{\um}           &  \0\0\0 3.5 &  \\
        Al\textsuperscript{13+}     & \SI{2}{\um}    &  \0\0\0 5.2 & \0 240 \\
        & \SI{5}{\um}           &  \0\0 12 \phantom{.}\0 &  \\
        Au\textsuperscript{51+}     & \SI{2}{\um}     &  \0\0 89 \phantom{.}\0 & \0\0 75 \\
        & \SI{5}{\um}           &  \0 190  \phantom{.}\0 & \\
        Au\textsuperscript{30+}     & \SI{5}{\um}    &  \0 140 \phantom{.}\0 & \0\0 31 \\
        \br
    \end{tabular}
    \caption{\label{conversion-material-compare}Efficiency of conversion of the laser pulse energy into
        the energy of all photons generated by the bremsstrahlung process $\etas{BS}$, and the inverse Compton 
        scattering process $\etas{ICS}$ for targets of different materials irradiated by a $a_0 = 86, 
        I=\SI{e22}{W/cm^2}$ laser pulse. For foils with $d > \SI{2}{\um}$, the conversion by inverse Compton 
        scattering does not depend on the target thickness, therefore the same values can be used for comparison of the 
        $d = \SI{2}{\um}$ and the $d = \SI{5}{\um}$ foil.}
\end{table}

\section{Conclusions}\label{conclusions}

We have studied the emission of gamma rays by inverse Compton scattering in interactions of a short
intense laser pulse with a thin foil target via 2D PIC simulations. The ICS process dominates 
over bremsstrahlung in low-Z targets already at a threshold intensity $I \approx \SI{3e21}{W/cm^2}$ under which no ICS 
generated gamma rays are seen at all. Spectra of the gamma rays produced in interactions with different driving pulse 
intensities show a linear dependence of the ICS produced gamma ray temperature on the intensity
$\Tgamma \sim a_0^2 \sim I$, at least in the studied intensity range $I = \num{3e21} - \SI{e23}{W/cm^2}$. 
As 
the ICS process takes place in front of the target in the evolving field of the laser pulse, the 
relation between the temperature of the electrons and that of the resulting gamma rays is provided as an empirical 
observation only.

The radiation is forward going with two lobes centred at approximately $\vartheta \approx \ang{\pm30}$. The 
angular distribution of the emission is dictated by the dynamics of the electrons in the field of the laser pulse in 
front of the target, thus for sufficiently thick $d \gtrsim \SI{2}{\um}$ targets, there is no change in its structure 
with increasing thickness. A simple theoretical model which assumes the movement of an electron in a planar 
standing wave formed in 
the front side by the interaction of the incoming and reflected parts of the laser pulse predicts the photon 
propagation angle 
$\vartheta = \ang{30}$ regardless of the laser pulse intensity. This is confirmed by the simulations up to 
$I \approx \SI{e22}{W/cm^2}$. As the intensity grows further, the propagation angle increases since the assumptions of 
the theoretical model break down due to hole boring. When the hole in the surface is sufficiently deep, the electrons
injected from its sides meet the laser pulse in a different phase, and travel along a different trajectory before being
reinjected near the centre of the hole. Moreover, when the hole's depth is comparable to the laser pulse wavelength
$\lambda = \SI{1}{\um}$, the combined field of the incoming and the reflected parts of the laser pulse cannot be
adequately described by that of a planar standing wave which would form in front of a flat surface.
Efficiency of conversion of the driving laser pulse energy into that of the gamma rays generated by ICS 
shows super-linear scaling with intensity $\etas{ICS} \sim I^{3/2}$ in the studied intensity range.

Comparing the results to our previous work, where we show that targets made of materials with a higher atomic number, 
while exhibiting a lower absorption, still show a significant increase of 
gamma ray production by bremsstrahlung \cite{VyskocilSimulationsbremsstrahlungemission2018}, we see that the 
lower absorption 
also affects the ICS process which does not directly depend on the atomic number. Lower-Z 
targets give out much more ICS gamma rays with a crossing point being a $d=\SI{2}{\um}$ thick Au\textsuperscript{51+} 
target irradiated by a $I=\SI{e22}{W/cm^2}$ laser pulse, for which the two processes exhibit roughly the same 
conversion efficiency.

\ack
We would like to than Mariana Kecov\'{a} from the ELI Virtual Beamline team for the visualization of 
electron 
trajectories in \fref{mari-tracks}. The results of this work were obtained under Project LQ1606 with the 
financial 
support of the Ministry of Education,
Youth and Sports as part of targeted support from the National Programme of Sustainability II. Supported by the
project ELITAS (CZ.02.1.01/0.0/0.0/16 013/0001793) and the project HiFI (CZ.02.1.01/0.0/0.0/15 003/0000449) from the
European Regional Development Fund. Additional funding was obtained under GAČR project 18-09560S, and the
SGS16/248/OHK4/3T/14 grant. The EPOCH code used in this research was developed under UK Engineering and Physics Sciences
Research Council grants EP/G054940/1, EP/G055165/1 and EP/G056803/1. Simulations were performed at the ECLIPSE cluster
at ELI-Beamlines, supported from the aforementioned ELI related grants, and the Salomon cluster at IT4Innovations, 
supported by the Ministry of Education, Youth and Sports from the Large Infrastructures for Research, Experimental 
Development and Innovations project ``IT4Innovations National Supercomputing Center -- LM2015070''.

\section*{References}
\bibliography{article.bib}

\providecommand{\newblock}{}
\begin{thebibliography}{10}
\expandafter\ifx\csname url\endcsname\relax
  \def\url#1{{\tt #1}}\fi
\expandafter\ifx\csname urlprefix\endcsname\relax\def\urlprefix{URL }\fi
\providecommand{\eprint}[2][]{\url{#2}}

\bibitem{WeberP3installationhighenergy2017}
Weber S, Bechet S, Borneis S, Brabec L, Bu{\v c}ka M, {Chacon-Golcher} E,
  Ciappina M, DeMarco M, Fajstavr A, Falk K, Garcia E~R, Grosz J, Gu Y~J,
  Hernandez J~C, Holec M, Jane{\v c}ka P, Janta{\v c} M, Jirka M, Kadlecova H,
  Khikhlukha D, Klimo O, Korn G, Kramer D, Kumar D, Lastovi{\v c}ka T,
  Lutoslawski P, Morejon L, Ol{\v s}ovcov{\'a} V, Rajdl M, Renner O, Rus B,
  Singh S, {\v S}mid M, Sokol M, Versaci R, Vr{\'a}na R, Vranic M, Vysko{\v
  c}il J, Wolf A and Yu Q 2017 {\em Matter and Radiation at Extremes\/} {\bf 2}
  149--176 ISSN 2468-080X

\bibitem{HernandezGomezVulcan10PW2010}
{Hernandez-Gomez} C, Blake S~P, Chekhlov O, Clarke R~J, Dunne A~M, Galimberti
  M, Hancock S, Heathcote R, {P Holligan}, Lyachev A, Matousek P, Musgrave I~O,
  Neely D, Norreys P~A, Ross I, Tang Y, Winstone T~B, Wyborn B~E and Collier J
  2010 {\em Journal of Physics: Conference Series\/} {\bf 244} 032006 ISSN
  1742-6596

\bibitem{ZouDesigncurrentprogress2015}
Zou J~P, Blanc C~L, Papadopoulos D~N, Ch{\'e}riaux G, Georges P, Mennerat G,
  Druon F, Lecherbourg L, Pellegrina A, Ramirez P, Giambruno F, Fr{\'e}neaux A,
  Leconte F, Badarau D, Boudenne J~M, Fournet D, Valloton T, Paillard J~L,
  Veray J~L, Pina M, Monot P, Chambaret J~P, Martin P, Mathieu F, Audebert P
  and Amiranoff F 2015 {\em High Power Laser Science and Engineering\/} {\bf 3}
  ISSN 2095-4719, 2052-3289

\bibitem{DansonPetawattclasslasers2015}
Danson C, Hillier D, Hopps N and Neely D 2015 {\em High Power Laser Science and
  Engineering\/} {\bf 3} ISSN 2095-4719, 2052-3289

\bibitem{BernsteinRayProductionLaser1970}
Bernstein M~J and Comisar G~G 1970 {\em Journal of Applied Physics\/} {\bf 41}
  729--733 ISSN 0021-8979

\bibitem{RitusQuantumeffectsinteraction1985}
Ritus V~I 1985 {\em Journal of Soviet Laser Research\/} {\bf 6} 497--617 ISSN
  0270-2010, 1573-8760

\bibitem{LauNonlinearThomsonscattering2003}
Lau Y~Y, He F, Umstadter D~P and Kowalczyk R 2003 {\em Physics of Plasmas\/}
  {\bf 10} 2155 ISSN 1070664X

\bibitem{NakamuraHighPowergRayFlash2012}
Nakamura T, Koga J~K, Esirkepov T~Z, Kando M, Korn G and Bulanov S~V 2012 {\em
  Physical Review Letters\/} {\bf 108} ISSN 0031-9007, 1079-7114

\bibitem{GuGammaphotonselectronpositron2019}
Gu Y~J, Jirka M, Klimo O and Weber S 2019 {\em Matter and Radiation at
  Extremes\/} {\bf 4} 064403 ISSN 2468-2047

\bibitem{BellPossibilityProlificPair2008}
Bell A~R and Kirk J~G 2008 {\em Physical Review Letters\/} {\bf 101} ISSN
  0031-9007, 1079-7114

\bibitem{BulanovElectromagneticcascadehighenergy2013}
Bulanov S~S, Schroeder C~B, Esarey E and Leemans W~P 2013 {\em Physical Review
  A\/} {\bf 87} ISSN 1050-2947, 1094-1622

\bibitem{DiPiazzaExtremelyhighintensitylaser2012}
Di~Piazza A, M{\"u}ller C, Hatsagortsyan K~Z and Keitel C~H 2012 {\em Reviews
  of Modern Physics\/} {\bf 84} 1177--1228

\bibitem{RidgersDenseElectronPositronPlasmas2012}
Ridgers C~P, Brady C~S, Duclous R, Kirk J~G, Bennett K, Arber T~D, Robinson
  A~P~L and Bell A~R 2012 {\em Physical Review Letters\/} {\bf 108} ISSN
  0031-9007, 1079-7114

\bibitem{BradyLaserAbsorptionRelativistically2012}
Brady C~S, Ridgers C~P, Arber T~D, Bell A~R and Kirk J~G 2012 {\em Physical
  Review Letters\/} {\bf 109} 245006

\bibitem{ZhidkovRadiationDampingEffects2002}
Zhidkov A, Koga J, Sasaki A and Uesaka M 2002 {\em Physical Review Letters\/}
  {\bf 88} 185002

\bibitem{TaPhuocAllopticalComptongammaray2012}
Ta~Phuoc K, Corde S, Thaury C, Malka V, Tafzi A, Goddet J~P, Shah R~C, Sebban S
  and Rousse A 2012 {\em Nature Photonics\/} {\bf 6} 308--311 ISSN 1749-4885,
  1749-4893

\bibitem{EnglertSecondharmonicphotonsinteraction1983}
Englert T~J and Rinehart E~A 1983 {\em Physical Review A\/} {\bf 28} 1539--1545

\bibitem{BulaObservationNonlinearEffects1996}
Bula C, McDonald K~T, Prebys E~J, Bamber C, Boege S, Kotseroglou T, Melissinos
  A~C, Meyerhofer D~D, Ragg W, Burke D~L, Field R~C, {Horton-Smith} G, Odian
  A~C, Spencer J~E, Walz D, Berridge S~C, Bugg W~M, Shmakov K and Weidemann A~W
  1996 {\em Physical Review Letters\/} {\bf 76} 3116--3119

\bibitem{ChenExperimentalobservationrelativistic1998}
Chen S~Y, Maksimchuk A and Umstadter D 1998 {\em Nature\/} {\bf 396} 653--655
  ISSN 1476-4687

\bibitem{SchwoererThomsonBackscatteredRaysLaserAccelerated2006}
Schwoerer H, Liesfeld B, Schlenvoigt H~P, Amthor K~U and Sauerbrey R 2006 {\em
  Physical Review Letters\/} {\bf 96} ISSN 0031-9007, 1079-7114

\bibitem{MalkaRelativisticelectrongeneration2002}
Malka G, Aleonard M~M, Chemin J~F, Claverie G, Harston M~R, Scheurer J~N,
  Tikhonchuk V, Fritzler S, Malka V, Balcou P, Grillon G, Moustaizis S,
  Notebaert L, Lefebvre E and Cochet N 2002 {\em Physical Review E\/} {\bf 66}
  066402

\bibitem{ChenMeVEnergyRaysInverse2013}
Chen S, Powers N~D, Ghebregziabher I, Maharjan C~M, Liu C, Golovin G, Banerjee
  S, Zhang J, Cunningham N, Moorti A, Clarke S, Pozzi S and Umstadter D~P 2013
  {\em Physical Review Letters\/} {\bf 110} ISSN 0031-9007, 1079-7114

\bibitem{SarriUltrahighBrillianceMultiMeV2014}
Sarri G, Corvan D~J, Schumaker W, Cole J~M, Di~Piazza A, Ahmed H, Harvey C,
  Keitel C~H, Krushelnick K, Mangles S~P~D, Najmudin Z, Symes D, Thomas A~G~R,
  Yeung M, Zhao Z and Zepf M 2014 {\em Physical Review Letters\/} {\bf 113}
  224801

\bibitem{YanHighordermultiphotonThomson2017}
Yan W, Fruhling C, Golovin G, Haden D, Luo J, Zhang P, Zhao B, Zhang J, Liu C,
  Chen M, Chen S, Banerjee S and Umstadter D 2017 {\em Nature Photonics\/} {\bf
  11} 514--520 ISSN 1749-4893

\bibitem{ColeExperimentalEvidenceRadiation2018}
Cole J~M, Behm K~T, Gerstmayr E, Blackburn T~G, Wood J~C, Baird C~D, Duff M~J,
  Harvey C, Ilderton A, Joglekar A~S, Krushelnick K, Kuschel S, Marklund M,
  McKenna P, Murphy C~D, Poder K, Ridgers C~P, Samarin G~M, Sarri G, Symes D~R,
  Thomas A~G~R, Warwick J, Zepf M, Najmudin Z and Mangles S~P~D 2018 {\em
  Physical Review X\/} {\bf 8} ISSN 2160-3308

\bibitem{PoderExperimentalSignaturesQuantum2018}
Poder K, Tamburini M, Sarri G, Di~Piazza A, Kuschel S, Baird C~D, Behm K,
  Bohlen S, Cole J~M, Corvan D~J, Duff M, Gerstmayr E, Keitel C~H, Krushelnick
  K, Mangles S~P~D, McKenna P, Murphy C~D, Najmudin Z, Ridgers C~P, Samarin
  G~M, Symes D~R, Thomas A~G~R, Warwick J and Zepf M 2018 {\em Physical Review
  X\/} {\bf 8} 031004

\bibitem{MalkaExperimentalConfirmationPonderomotiveForce1996}
Malka G and Miquel J~L 1996 {\em Physical Review Letters\/} {\bf 77} 75--78

\bibitem{PukhovThreeDimensionalSimulationsIon2001}
Pukhov A 2001 {\em Physical Review Letters\/} {\bf 86} 3562--3565

\bibitem{ArberContemporaryparticleincellapproach2015}
Arber T~D, Bennett K, Brady C~S, {Lawrence-Douglas} A, Ramsay M~G, Sircombe
  N~J, Gillies P, Evans R~G, {H Schmitz}, Bell A~R and Ridgers C~P 2015 {\em
  Plasma Physics and Controlled Fusion\/} {\bf 57} 113001 ISSN 0741-3335

\bibitem{VyskocilSimulationsbremsstrahlungemission2018}
Vysko{\v c}il J, Klimo O and Weber S 2018 {\em Plasma Physics and Controlled
  Fusion\/} {\bf 60} 054013 ISSN 0741-3335, 1361-6587

\bibitem{KirkPairproductioncounterpropagating2009}
Kirk J~G, Bell A~R and Arka I 2009 {\em Plasma Physics and Controlled Fusion\/}
  {\bf 51} 085008

\bibitem{SauterUeberVerhaltenElektrons1931}
Sauter F 1931 {\em Zeitschrift f{\"u}r Physik\/} {\bf 69} 742--764 ISSN
  0044-3328

\bibitem{SchwingerGaugeInvarianceVacuum1951}
Schwinger J 1951 {\em Physical Review\/} {\bf 82} 664--679 ISSN 0031-899X

\bibitem{ShenEnergyStragglingRadiation1972}
Shen C~S and White D 1972 {\em Physical Review Letters\/} {\bf 28} 455--459

\bibitem{NikishovQuantumprocessesfield1964}
Nikishov A and Ritus V 1964 {\em Sov. Phys. JETP\/} {\bf 19} 529--541

\bibitem{WilksAbsorptionultraintenselaser1992}
Wilks S~C, Kruer W~L, Tabak M and Langdon A~B 1992 {\em Physical Review
  Letters\/} {\bf 69} 1383--1386

\bibitem{YeeNumericalsolutioninitial1966a}
Yee K 1966 {\em IEEE Transactions on Antennas and Propagation\/} {\bf 14}
  302--307 ISSN 0018-926X

\bibitem{Boris1970}
Boris J~P 1970 {\em Proceeding of Fourth Conference on Numerical Simulations of
  Plasmas\/}

\bibitem{DuclousMonteCarlocalculations2010}
Duclous R, Kirk J~G and Bell A~R 2010 {\em Plasma Physics and Controlled
  Fusion\/} {\bf 53} 015009

\bibitem{JiEnergypartitiongray2014}
Ji L~L, Pukhov A, Nerush E~N, Kostyukov I~Y, Shen B~F and Akli K~U 2014 {\em
  Physics of Plasmas\/} {\bf 21} 023109 ISSN 1070-664X

\bibitem{ChangUltraintenselaserabsorption2017}
Chang H~X, Qiao B, Zhang Y~X, Xu Z, Yao W~P, Zhou C~T and He X~T 2017 {\em
  Physics of Plasmas\/} {\bf 24} 043111 ISSN 1070-664X

\bibitem{RobinsonRelativisticallycorrectholeboring2009}
Robinson A~P~L, Gibbon P, Zepf M, Kar S, Evans R~G and Bellei C 2009 {\em
  Plasma Physics and Controlled Fusion\/} {\bf 51} 024004 ISSN 0741-3335,
  1361-6587

\bibitem{BradyGammarayemissioncritical2013}
Brady C~S, Ridgers C~P, Arber T~D and Bell A~R 2013 {\em Plasma Physics and
  Controlled Fusion\/} {\bf 55} 124016 ISSN 0741-3335

\bibitem{BradySynchrotronradiationpair2014}
Brady C~S, Ridgers C~P, Arber T~D and Bell A~R 2014 {\em Physics of Plasmas\/}
  {\bf 21} 033108 ISSN 1070-664X

\end{thebibliography}

\end{document}